\documentclass[12pt,preprint]{aastex}

\usepackage[dvips]{color} 

\shorttitle{Sequential Star Formation in NGC~604}
\shortauthors{Rie Miura, et al.}

\begin{document}


\title{Aperture Synthesis Observations of CO, HCN, and 89\,GHz Continuum Emission toward NGC~604 in M~33: Sequential Star Formation Induced by Supergiant H\,{\sc ii} region}


\author{Rie Miura\altaffilmark{1,2}, Sachiko  K. Okumura\altaffilmark{1,2}, Tomoka
Tosaki\altaffilmark{3}, Yoichi Tamura\altaffilmark{4,2}, Yasutaka Kurono\altaffilmark{1}, Nario Kuno\altaffilmark{4}, Kouichiro
Nakanishi\altaffilmark{1}, Seiichi Sakamoto\altaffilmark{5}, Takashi Hasegawa\altaffilmark{6} \and  Ryohei Kawabe\altaffilmark{4}}
\affil{$^1$National Astronomical Observatory of Japan, 2-21-1 Osawa, Mitaka, Tokyo, 181-8588, Japan}
\email{rie.miura@nao.ac.jp}
\affil{$^2$Department of Astronomy, The University of Tokyo,
Hongo, Bunkyo-ku, Tokyo, 133-0033, Japan}
\affil{$^3$Joetsu University of Education, Yamayashiki-machi, Joetsu, Niigata, 943-8512, Japan} 
\affil{$^4$Nobeyama Radio Observatory, Minamimaki, Minamisaku, Nagano,
384-1805, Japan}
\affil{$^5$Institute of Space and Astronautical Science, 3-1-1 Yoshinodai, Chuo-ku, Sagamihara-shi, Kanagawa 252-5210, Japan}
\affil{$^6$Gunma Astronomical Observatory, Nakayama, Takayama,
  Agatsuma, Gunma, 377-0702, Japan}


\begin{abstract}
We present the results from new Nobeyama Millimeter Array observations of CO(1--0), HCN(1--0), and 89-GHz continuum emissions toward NGC~604, known as the supergiant H~{\sc ii} region in a nearby galaxy M~33. Our high spatial resolution images ($4 \farcs 2 \times 2 \farcs 6$, corresponding to 17 pc $\times$ 11 pc physical size) of CO emission allowed us to uncover ten individual molecular clouds that have masses of (0.8 -- 7.4) $\times 10^5 M_{\sun}$ and sizes of 5 -- 29 pc, comparable to those of typical Galactic giant molecular clouds (GMCs). Moreover, we detected for the first time HCN emission in the two most massive clouds and 89\,GHz continuum emission at the rims of the ``H$\alpha$ shells''. The HCN and 89 \,GHz continuum emission show offset from the CO peak and are distributed to the direction of the central cluster. Three out of ten CO clouds are well correlated with the H$\alpha$ shells both in spatial and velocity domains, implying an interaction between molecular gas and the expanding H~{\sc ii} region.  The CO clouds show varieties in star formation efficiencies (SFEs), which are estimated from the 89-GHz and combination of H$\alpha$ and {\it Spitzer} 24-$\micron$ data. Furthermore, we found that the SFEs decrease with increasing projected distance measured from the heart of the central OB star cluster in NGC~604, suggesting the radial changes in evolutionary stages of the molecular clouds in course of stellar cluster formation. Our results provide further support to the picture of sequential star formation in NGC604 initially proposed by \citet{2007ApJ...664L..27T} with the higher spatially resolved molecular clouds, in which an isotropic expansion of the H~{\sc ii} region pushes gases outward and accumulates them to consecutively form dense molecular clouds, and then induces massive star formations.
\end{abstract}





\section{INTRODUCTION}
The expansion of an H\,{\sc ii} region can trigger star formation in various ways \citep[see the review of][]{1998ASPC..148..150E}: direct compression of pre-existing clouds and globules due to high pressure of H\,{\sc ii} regions (``globule squeezing''), and gas accumulation and cloud collapse at the edges of H\,{\sc ii} regions and supernovae remnant (``collect and collapse''). 
The globule squeezing process works on the smaller scale of cometary globules, while the collect-and-collapse process on the larger scale of the H\,{\sc ii} region. 
Among above two processes, the collect and collapse process, first proposed by \citet{1977ApJ...214..725E}, is particularly interesting as it allows the formation of massive fragments to form massive stars or clusters \citep{1994MNRAS.268..291W}. As an H{\sc ii} region expands with supersonic velocity, the ambient medium is photoionized and dynamically swept up into shocked layer, which may become gravitationally unstable and collapse to filaments, and then into one or more dense cores to form new generation stars. 
Many multi-wavelength studies for the collect and collapse process have been performed in our Galaxy and neighboring galaxies \citep[see examples in][]{1998ASPC..148..150E}. 
From {\it JHK} photometry, The stellar clusters of the different age are seen spatially resolved in an H\,{\sc ii} region, and moreover, embedded younger massive clusters are found to lie at the peripheries of H\,{\sc ii} regions \citep[e.g.,][]{2005A&A...433..565D,2006A&A...446..171Z,2008A&A...482..585D}. 
These clusters are possible second-generation clusters whose formation has been triggered by the expansion of H\,{\sc ii} region \citep[``propagating star formation'';][]{1998ASPC..148..150E}.

\citet{1994AJ....108.1276M,1995AJ....109.2503M} and \citet{1996AJ....111.1252M} extended this process of propagating star formation in molecular clouds to a larger scale to explain the star formation process in Giant H\,{\sc ii} regions (GHRs). GHRs can easily spread over areas 10$^2$--10$^3$\,pc across, and play a vital role in determining the large-scale structures of their host galaxies. However, little is known about the possible connection between ambient interstellar matter and expansion of GHRs over such distances. This is because there is not such a giant H\,{\sc ii} region in our Galaxy, and thus toward external galaxies it has been difficult to perform high-resolution, high-sensitivity, and wide-field observations at once. 
Evidences of ``collect and collapse'' process on large-scale were most notably from the correspondence between the old H\,{\sc i} holes and young OB associations \citep{1986A&A...169...14B,1990A&A...229..362D}, showing the kinematic and spatial evidences of expansion of H\,{\sc i} and H$\alpha$ shells. \citet{1990A&A...229..362D} allow for the possibility that many such holes have been excavated by massive star formation activity and formed by the local collapse of H\,{\sc i} atomic gas to form giant molecular clouds (GMCs) which are the principal sites of star formation. In fact, the GMCs tend to be spatially clustered around the H\,{\sc i} holes \citep{2003ApJS..149..343E}.

The spatially resolved studies of these GMCs are necessary in order to understand the triggers of star formation around GHRs through the formation of dense molecular clouds which will be taken place in individual GMC. 
Recent progress of millimeter arrays provides wide-field, high-sensitivity, and high-resolution images of the molecular gas for relatively shorter observational time, which has potential to reveal clumps within GMCs and physical relation to ongoing star formation. 
Our objective is to study how extragalactic GMCs evolve to form stars through dense gas formation which might be affected by large-scale phenomena that is, expansion of giant H\,{\sc ii} region. 
In particular we focus on the evolution of GMCs by probing their physical nature such as their density structures and star formation activities, and their spatial distributions against the activities in galactic environments from the multi-wavelength observations. With this purpose, we embarked upon detailed studies toward nearby extragalactic GMCs in a bright supergiant H\,{\sc ii} region, NGC~604.

The giant H\,{\sc ii} region NGC~604 is located in northern arm in the spiral galaxy \objectname{M~33}. 
The proximity \citep[distance of 0.84 Mpc;][]{2001ApJ...553...47F} and favorable inclination of M33 \citep[${\it i}=51\arcdeg$;][]{2000MNRAS.311..441C} are sufficient to resolve gas components with little contamination along the line of sight and to understand the inner structure of GMCs \citep[i.e., typically 40\,pc in diameter;][]{1987ASSL..134...21S}. 
Because of its striking features, the giant H\,{\sc ii} region NGC~604 has been observed in a wide range of wavelengths; radio \citep{1992A&A...265..437V, 1992ApJ...385..512W,1999ApJ...514..188C,2007ApJ...664L..27T}, optical \citep{1993AJ....105.1400D,1996ApJ...456..174H,2000ApJ...541..720T,2004AJ....128.1196M}, and X-rays \citep{2008ApJ...685..919T}.
According to previous studies, the H$\alpha$ nebula has a core-halo structure extending up to a radius of 200 -- 400 pc \citep[][see also Figure \ref{fg:total}]{1980A&A....86..304M}. 
The H$\alpha$ core, containing a very young (3--5\,Myr) and rich (over 200) O-type stellar population, is surrounded by photo-ionized filaments and a section of shells \citep{1993AJ....105.1400D, 1996ASPC...98..399D, 1996ApJ...456..174H,  2000MNRAS.317...64G, 2000ApJ...541..720T, 2003AJ....125.3082B, 2009ApJ...699.1125R}.  
The energy of the stellar wind has been estimated to be enough to form the core of the H$\alpha$ nebula \citep[e.g.,][]{2000ApJ...541..720T, 2004AJ....128.1196M, 2008ApJ...685..919T}, on the other hand, SNRs as well as stellar winds would contribute to blow it up to the halo structure \citep[e.g.,][]{1996AJ....112..146Y}. As noted in Figure \ref{fg:total}, we call the shell surrounding the main cavity ``Shell~A" and the shell located to the south of Shell~A, ``Shell~B", which appears to be a blowout through the molecular tunnel over Shell~A \citep[][]{2000ApJ...541..720T}. 
OB and W-R stars are mostly distributed in Shell~A, while, a small number of OB stars in Shell~B \citep[see Figure 7 in][]{1996AJ....112..146Y}. Near-infrared $JHK$ photometric study of NGC~604 was recently reported in \citet{2010IAUS..266..391F}, the candidates of massive young stellar objects (MYSO) appear aligned with bright H$\alpha$ Shells A and B. An H\,{\sc i} hole surrounds around diffuse H$\alpha$ nebula and some GMCs are within 150\,pc around the H\,{\sc i} hole edge \citep{2003ApJS..149..343E,2009ApJ...700..545H}. 
The existence of five GMCs over the extent H$\alpha$ emission in the NGC~604 region was revealed by previous $^{12}$CO observations \citep{1992A&A...265..437V, 1992ApJ...385..512W, 2003ApJS..149..343E, 2003ApJ...599..258R}. 
The sensitivity and spatial resolution of these observations using the OVRO interferometer and the BIMA array were, however, not sufficient to resolve the inner structure of any of these GMC.

We present here the new results of the interferometric CO($J$=1--0), HCN, and 89\,GHz continuum observations toward the NGC~604 region. Our imaging achieved a higher spatial resolution than any other studies so far (10\,pc scale for $^{12}$CO emission). We observed the $J$=1--0 emission lines of $^{12}$CO and HCN, and adjacent 89\,GHz continuum emission. 
These three probes trace different components of the GMC; i.e., relatively diffuse molecular clouds (the critical density of $^{12}$CO line emission, $n_{\rm H_2}\sim10^2\,{\rm cm}^{-3}$), denser regions ($n_{\rm H_2}>10^4\,{\rm cm}^{-3}$), and active star-forming regions, respectively. Indeed, the HCN line emission has been known to trace denser regions in molecular clouds linked to star-formation activities rather than $^{12}$CO \citep[e.g.,][]{1992ApJ...387L..55S}. The 89\,GHz continuum is considered to be mainly dominated with free-free emission from ionized gas by young OB stars, and thus expected to trace massive star forming region. The HCN observation and millimeter-wavelength continuum emission toward NGC~604 are reported for the first time in this work.
Besides, this work provides the first example of extragalactic GMCs in spiral galaxies being observed on a size scale comparable to typical galactic GMCs.

The outline of this paper is as follows.
The observations and data reduction are summarized in $\S$ \ref{sec:2}.
In $\S$ \ref{sec:3} the general procedure for identifying clouds in the
survey and the results of each probe are described. 
In the discussion in $\S$ \ref{sec:4}, we quantify the star formation activities in the GMCs using our data and also 24~$\mu$m and H$\alpha$ data, and discuss the varieties of the physical states of the identified clouds.
We also discuss the relationship between the nature of these clouds and their evolutionary stage. Finally, we interpret our results in the context of a sequential star formation scenario, as initially proposed for NGC604 by \citet{2007ApJ...664L..27T} from the spatial relationship between the GMCs and H$\alpha$ nebula. 
The summary of this work is given in $\S$\ref{sec:5}.

\section{Observations and Data Reduction}
\label{sec:2}
\begin{figure*}
\center
\includegraphics*[scale=1.0,trim=150 50 15 0, clip]{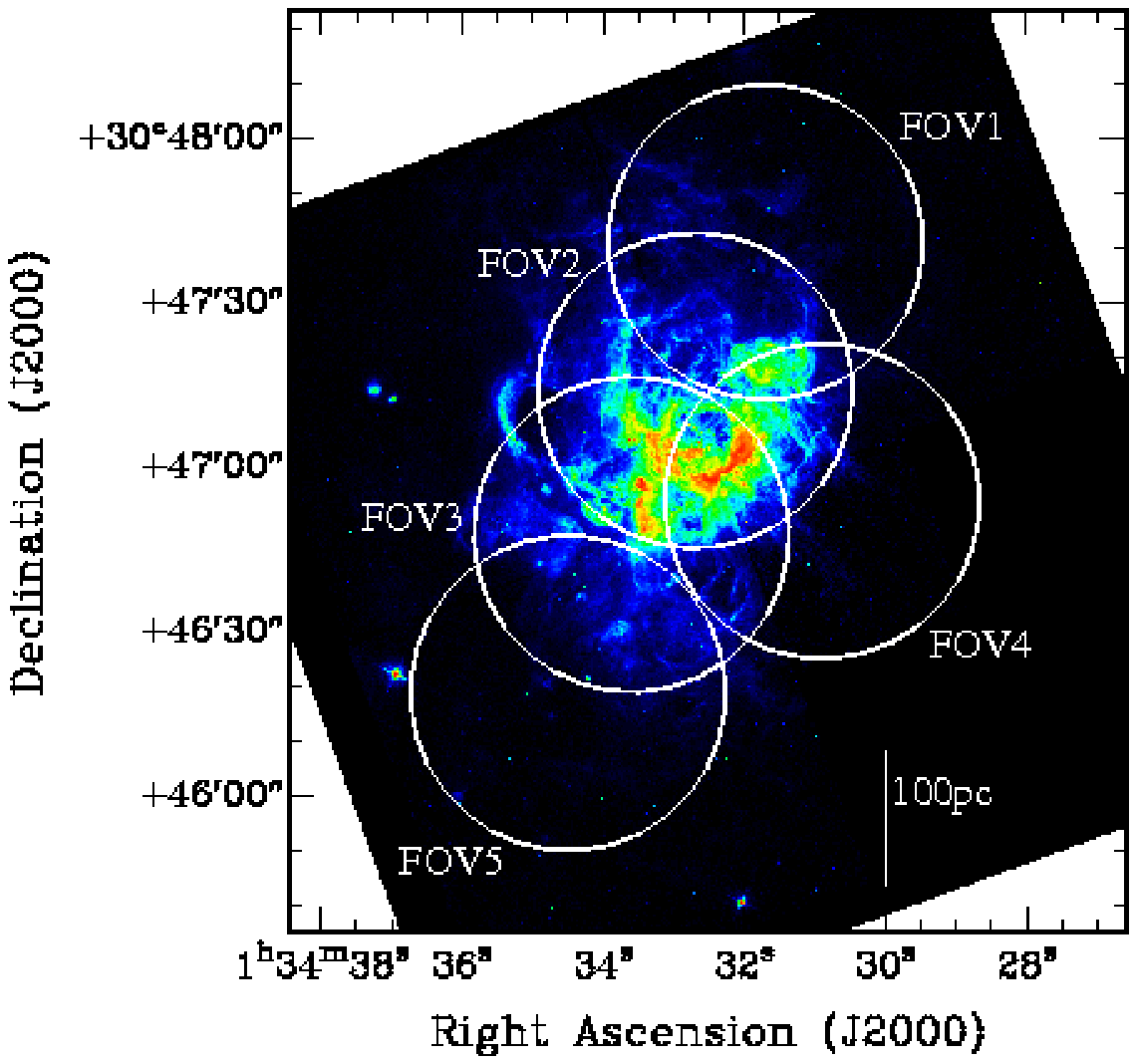}
\caption{Positions of the five fields of view (FoVs) observed in $^{12}$CO(1--0) overlaid on the distribution of the H$\alpha$ which is available in the HST archive. The archival HST WFPC2 image of NGC~604 from program No. 5237 (PI J. Westphal) is used here.
Each circle corresponds to the NMA primary beam (FWHM$=59\arcsec$) at 115\,GHz. The field names are also labeled. A bar on the right bottom in the figure represents linear scale of 100\,pc at the distance of M33. \label{fg:nmafov}}
\end{figure*}

We performed $^{12}{\rm CO}$(1--0), HCN(1--0), and 89\,GHz continuum observations toward \objectname{NGC~604} using the Nobeyama Millimeter Array (NMA) at Nobeyama Radio Observatory (NRO) from November 2004 to December 2006. 
We observed $^{12}$CO line emission toward the five fields of view (FoVs) with both C and D configurations of NMA (see Figure \ref{fg:nmafov}, Table \ref{tb:fovsummary}, and Table \ref{tb:nmaobservationalparameter}). 
The area of the five FoVs was about $110\arcsec\,\times\,145\arcsec$, corresponding to 450\,pc$\times$600\,pc covering the entire spatial extent of the optical nebulae. 
Besides, HCN(1--0) and 89\,GHz continuum emission were observed toward two out of the five FoVs to cover the two largest GMCs. Toward other three FoVs including smaller GMCs, we have not observed because it is expected to need more integration time to detect than toward the two largest GMCs.
The sizes of FoVs are represented by the full width at half-maximum (FWHM) of the primary beam at each frequency and they are given in Table \ref{tb:nmaobservationalparameter}.

We used DSB SIS receivers as the front-end, and the digital
spectro-correlator New-FX (1024 channels, 32\,MHz bandwidth) as the back-end for the $^{12}$CO and HCN observations.
The total bandwidth of the FX correlator corresponds to $83.2\,{\rm km\,s}^{-1}$
at the $^{12}$CO frequency, and $121.6\,{\rm km\,s}^{-1}$ at HCN, respectively.
The 89\,GHz continuum observations were performed simultaneously with the HCN observations using the Ultra-Wide-Band Correlator \citep[UWBC;][]{2000PASJ...52..393O} which has 128 channels and a bandwidth of 1024\,MHz. 
The continuum emission data were made by merging the frequency information in the IF signals from lower sideband (LSB) of the UWBC. 
The frequency ranges containing line emissions of HCN and HCO$^+$ (89.188526\,GHz) were removed from the original visibility data. The effective bandwidth of line-free channels is 768\,MHz.

The bandpass calibration was done using the quasar 3C~454.3. The radio sources B0234+285 and B0133+476 were observed every 20 minutes as phase calibrators.
The flux scales of these calibrators were determined by comparison
with Uranus and Neptune. 
The uncertainty in absolute flux scale is expected to be less than 15\,\%.

The interferometric data were reduced using the NRO software package UVPROC-II  \citep{1997ASPC..125...50T}. 
The final mosaiced image was made by combining the visibility data of all the available FoVs with the software package MIRIAD (using the task {\sf invert} to make the synthesis image, and {\sf mossdi} to CLEAN the image for the line data or {\sf mfclean} for the continuum data).
All these maps were made with natural {\it uv} weighting.
The synthesized beam sizes of the $^{12}$CO images for each FoV are given in Table \ref{tb:fovsummary}, while those of the HCN and 89\,GHz continuum are in Table \ref{tb:nmaobservationalparameter}. 
Note that the spatial resolutions in the partially overlapped areas between FoVs are the average of those in the overlapping FoVs.

The final 3-dimensional data cube of the $^{12}$CO and the HCN data has a frequency resolution of 1\,MHz (corresponding to $2.6\,{\rm km\,s}^{-1}$ at 115 GHz, $3.4\,{\rm km\,s}^{-1}$ at 89\,GHz, respectively) and the typical noise level of them are given in Table \ref{tb:fovsummary} and Table \ref{tb:nmaobservationalparameter}. The noise level of the 89\,GHz continuum data is also given in Table \ref{tb:nmaobservationalparameter}. 
The noise level was measured over the emission-free area. 
The final flux densities and noise levels for the sources were estimated after correcting for the primary beam attenuation.
Note that the noise levels rise by a factor of 1.5 at the edges of each FoV due to the primary beam correction.
Since the {\it uv} coverage of each pointing is not identical, the synthesized beam patterns differ among pointings. 
We briefly summarize the observational parameters in Table \ref{tb:fovsummary} and Table \ref{tb:nmaobservationalparameter} for the interferometric observations.

It is worth noting that interferometric observations do not have the sensitivity to large scale features (i.e., low spatial frequencies) of size larger than $\sim \lambda /B_{\rm min}$, where $\lambda$ is the observing wavelength and $B_{\rm min}$ is the minimum projected baseline. For our observations, $B_{\rm min}$ is 8--9\,k$\lambda$ at 115\,GHz, and 4 k$\lambda$ at 89\,GHz. 
Thus emission structures extended over spatial scales more than 23$\arcsec$--26$\arcsec$ (93--105$\,{\rm pc})$ in 115\,GHz observation, and $52\arcsec(\sim 211\,{\rm pc})$ in 89\,GHz observation are resolved out in the interferometric images. 
In order to estimate the missing flux in the CO observations, we compared the total integrated fluxes for each FoV in our NMA maps to the single-dish maps obtained using the 45-m telescope at NRO (R. Miura et al., 2010, in preparation). 
The fluxes were measured within a box of $42\arcsec\,\times\,42\arcsec$ at the center of each FoV. The interferometric to single-dish flux ratios are given in column 8 of Table \ref{tb:fovsummary}. The estimated ratios of FoV2 and FoV4 are relatively lower than other three FoVs, less than 30\%, because there is little molecular gas that detected in interferometric CO observation although the single-dish maps in FoV2 and FoV4 includes the envelope of the three largest GMCs which are located at the center of FoV1, FoV3, and FoV5, respectively.
The averaged ratio over the five FoVs is 0.46 $\pm$ 0.17, suggesting that half of the CO emission is resolved out in our interferometer observations. 





\section{Results}
\label{sec:3}
\subsection{$^{12}$CO(1--0)}
\label{sec:co}
\begin{figure*}
\includegraphics*[scale=1.0,trim=0 0 0 0, clip]{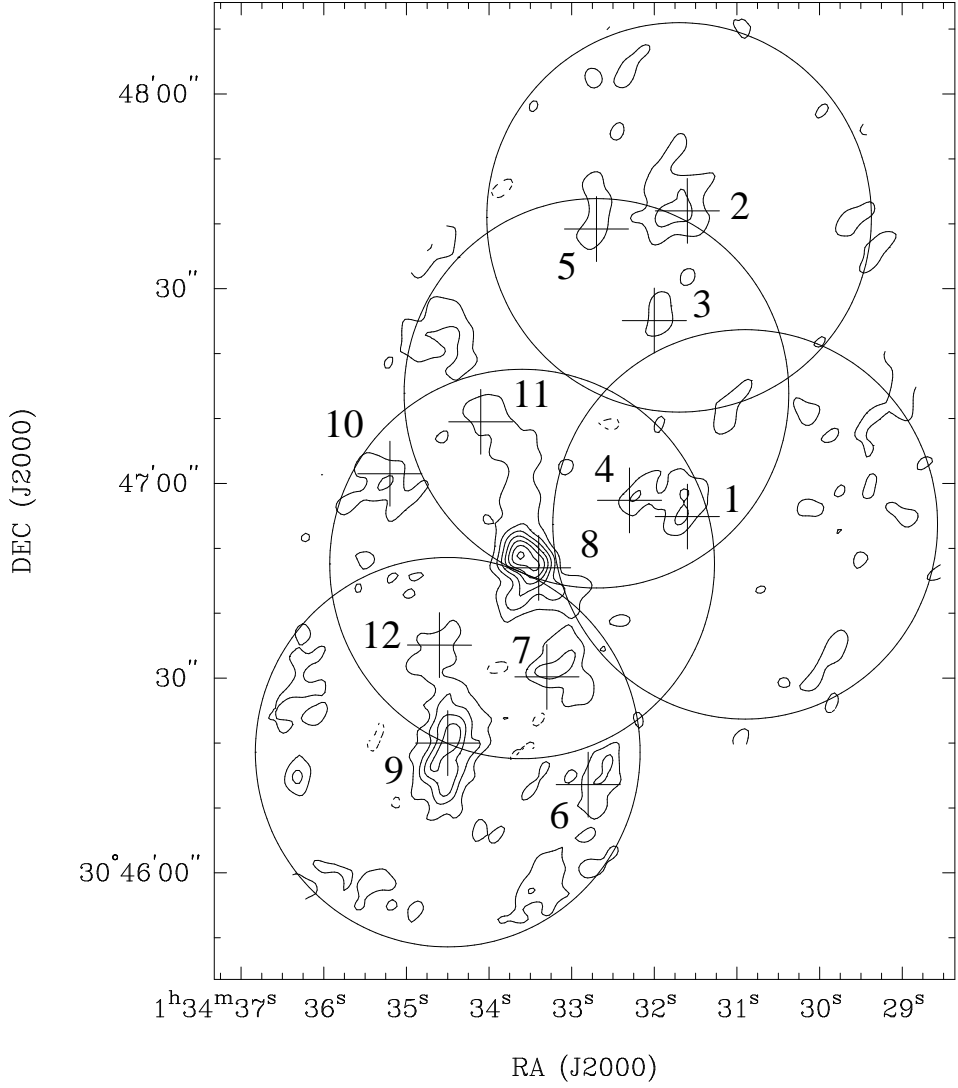}
\caption{Total integrated intensity map of $^{12}$CO over five FoVs. Primary beam correction was applied to this image. Contour levels are -3, 3, 6, 9, 12, 15, 18, and 21\,$\sigma_{\rm rms}$, where 1\,$\sigma_{\rm rms}=1.8 \times 10^3\,{\rm MJy\,sr}^{-1}\,{\rm km\,s}^{-1}$. To convert to ${\rm Jy\,sr}^{-1}$, the conversion factor 2.8 $10^{-10}$ sr\,beam$^{-1}$ was used.
The crosses represent the positions of the 12 identified clouds and their names are also labeled. The circles represent five FoVs that we observed in CO line emission. The parts of the emission outside the FoVs are masked out. 
\label{fg:comoment}}
\end{figure*}

The CO integrated intensity map in the velocity range $V_{\rm LSR}=-211\,{\rm\,km\,s^{-1}}$ to $-256\,{\rm\,km\,s^{-1}}$ is shown in Figure \ref{fg:comoment}. 
We identified CO clouds using the following criteria;
cloud emission must be more than three times stronger than the rms noise level ($\sigma_{\rm rms}$) and at least two times wider in velocity than our velocity resolution, and the cloud sizes must be larger than the equivalent area of the synthesized beam size. Ten clouds were identified with these above criteria. Their physical properties were obtained as described below. 

The cloud size $\bar{D}$ is defined as the average of the deconvolved FWHM diameters in right ascension and declination. 
The typical error of $\bar{D}$ was estimated as an average of the spatial resolution in the individual pointings. The velocity FWHM ($V_{\rm FWHM}$) was derived from Gaussian fits to emission line. Virial masses ($M_{\rm vir}$) and molecular cloud masses ($M_{\rm mol}$) were calculated using the procedures described in \citet[][and references therein]{1990ApJ...363..435W}. They were derived using the equations, $M_{\rm vir}=99 {D_{\rm pc}}{\rm V}^2_{\rm FWHM}\,{\rm M}_{\sun}$ and
$M_{\rm mol}=1.61 \times 10^4 d^2_{\rm Mpc} S_{\rm CO}\,{\rm M}_{\sun}$, where 
$D_{\rm pc}$ is the cloud size in parsec, $d_{\rm Mpc}$ is the distance to M~33 in Mpc, and $S_{\rm CO}$ is the CO flux density in ${\rm Jy\,km\,s^{-1}}$. We used 1.4 times the deconvolved diameter of the cloud in calculating the virial mass, since this diameter was found empirically to contain most of the flux of the cloud \citep{1990ApJ...363..435W}.
Here we assume the galactic value of the conversion factor from CO flux densities and H$_2$ column densities, $(3\pm1)\times10^{20}\,$cm$^{-2}$/(K\,km\,s$^{-1}$) \citep{1988A&A...207....1S}, and a factor of 1.36 has been included for the contribution of helium.

Table \ref{tb:ngc604cloudproperties} summarizes the properties of the ten clouds we identified. The positions of the ten clouds are shown in Figure \ref{fg:comoment}. The cloud sizes range from less than 5\,pc to 29\,pc and their masses are in the range (0.8 -- 7.4)$\times$10$^5\,$M$_{\sun}$, which are comparable to those of typical GMCs in the Milky Way \citep[e.g.,][]{1987ASSL..134...21S}. Six clouds (NMA-1, NMA-3, NMA-4, NMA-5, NMA-6, and NMA-10 in Table \ref{tb:ngc604cloudproperties}) were newly identified with our observations. The rest were already reported by the previous papers \citep{1992ApJ...385..512W, 2003ApJS..149..343E}.

We would like to mention two other clouds, NMA-11 and NMA-12. Even though their features do not satisfy our criteria for cloud identification, they were detected over \,3$\,\sigma$ but only one velocity channel, they were also detected in our single dish image with the NRO 45-m telescope (R. Miura et al., 2010, in preparation). One source NMA-11 is located to the north of NMA-8, while the other source NMA-12 is to the north of NMA-9. The properties of these two clouds are also given in Table \ref{tb:ngc604cloudproperties}.

\subsection{HCN} 
\label{sec:3.2}
\begin{figure*}
\begin{center}
\epsscale{1.0}
\plottwo{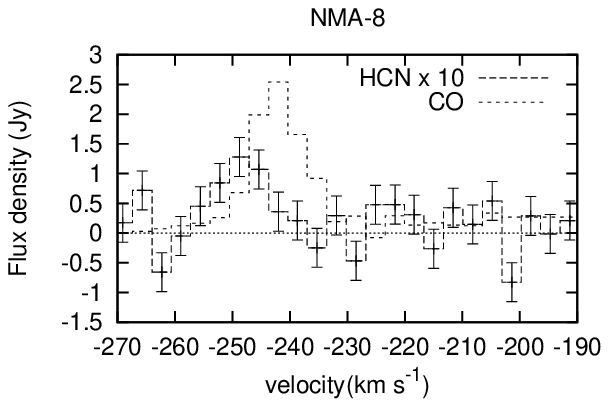}{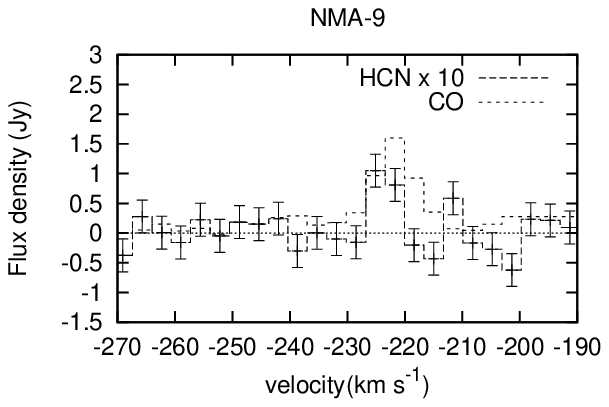}
\caption{The spectra of the two HCN components. The {\it left} panel is one component associated with the CO molecular cloud NMA-8, while the {\it right} panel is the other associated with NMA-9. The CO spectra for NMA-8 and NMA-9 are also shown, which are measured in the same region on the convolved image with the same synthesized beam and with the same velocity resolution as the HCN image.\label{fg:HCNspectrum}}
\end{center}
\end{figure*}

\begin{figure*}
\begin{center}
\includegraphics[scale=0.50, trim=80 0 0 0, clip]{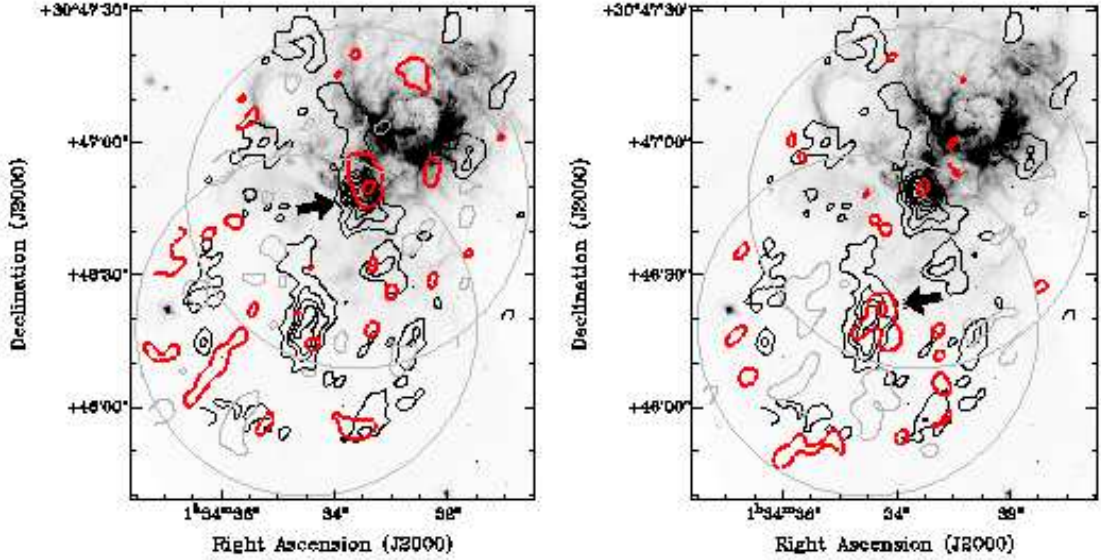}
\caption{Total integrated intensity map of HCN (red contours) and CO (black contour) overlaid on the H$\alpha$ image (gray scale) for NMA-8 ({\it left}) and NMA-9 ({\it right}). The integrated velocity ranges are from V$_{\rm LSR}=-245$\,km\,s$^{-1}$ to $-252$\,km\,s$^{-1}$ for NMA-8, and from V$_{\rm LSR}=-222$\,km\,s$^{-1}$ to $-225$\,km\,s$^{-1}$ for NMA-9. The contour levels for HCN map are $-2.5, 2.5, 5\,\sigma_{\rm rms}$, where 1\,$\sigma_{\rm rms}=1.5\times\,10^2\,{\rm MJy\,sr}^{-1}\,{\rm km\,s}^{-1}$. The conversion factor from Jy\,beam$^{-1}$ to Jy\,sr$^{-1}$ is $9.3 \times 10^{-10}$ sr\,beam$^{-1}$ for the $6 \farcs 6 \times 5 \farcs 3$ resolution data. The negative contours are drawn as gray contour. The black contour levels for CO are 3, 6, 9, 12, 15, 18, and 21\,$\sigma_{\rm rms}$. The gray circles with the diameter of 77\arcsec are two FoVs that we observed in HCN and 89\,GHz continuum emission. The black arrows indicate the region where HCN emission was detected over two channels and with $> 3\,\sigma$. The parts of the emission outside the FoVs are masked out. Primary beam correction was applied to these images. \label{fg:hcnmoment}}
\end{center}
\end{figure*}

Two HCN emission components were marginally detected more than three times stronger than the rms noise level and over two velocity channels (each has $3.4\,{\rm\,km\,s^{-1}}$). One is at $(01^{\rm h}34^{\rm m}33\fs4$, $+30\arcdeg46\arcmin50\farcs5)$ at the velocity of $-245\,{\rm km\,s^{-1}}$ and the other at $(01^{\rm h}34^{\rm m}34\fs3$, $+30\arcdeg46\arcmin21\farcs0)$ at the velocity of $-225\,{\rm km\,s^{-1}}$. 
Figure \ref{fg:hcnmoment} shows the HCN integrated intensity map overlaid on the CO map, which were integrated over the velocity range where the emission was seen ($>\,3\,\sigma_{\rm rms}$). These two HCN clouds seem to be associated with CO clouds, NMA-8 and NMA-9. The peaks of both HCN clouds show offset from those of CO clouds, and moreover, both of HCN clouds are located on the northern side of each CO cloud, to the direction of center of H\,{\sc ii} region. 
To check this association we have compared the velocities of the CO clouds to those of the HCN clouds. The HCN and CO spectra of these two components at the velocity resolution of $3.4\,{\rm\,km\,s^{-1}}$ are shown in Figure \ref{fg:HCNspectrum}. Each spectrum was measured over the area where the emission was seen ($>\,3\,\sigma_{\rm rms}$). 
The CO images were convolved to the same spatial and velocity resolution of the HCN images and the CO spectra were measured in the same region as the HCN spectra. The peak of the HCN spectra is slightly blue-shifted compared to that of the CO spectra in the same spatial region, which are found in both profiles of NMA-8 and NMA-9.
The bluer velocity corresponds to the direction of downstream by considering the direction of galactic circular rotation (from south to north in Figure \ref{fg:hcnmoment}) in this region, and hence these two HCN components seem to be associated with the blue components of CO clouds, which are located on the downstream (northern) side.

This suggests that the components with different velocities have different flux ratio of HCN to CO. The flux ratio of HCN to CO is often used as a density probe \citep[e.g.,][]{1997ApJ...478..162H}. 
We convolved the CO images with a Gaussian function of the same beam size as that of the HCN images for comparison between the total flux densities in the CO and HCN images at the same spatial scale. 
We obtained each total flux density by integrating over the extent of the emission components with more than $3\,\sigma_{\rm rms}$. 
The flux ratios of HCN to CO ($S_{\rm HCN}/S^\ast_{\rm CO}$ \footnote{We define the CO flux densities in the convolved images as $S^\ast_{\rm CO}$.}) were estimated to be $0.024\,\pm\,0.007$ for NMA-8 and $0.018\,\pm\,0.007$ for NMA-9. For the CO clouds where we did not detect any HCN emission (NMA-1, NMA-4, NMA-6, NMA-7, and NMA-10), we set upper limits on the line ratio using the detection limit of $3\,\sigma_{\rm rms}$. The upper limits of $S_{\rm HCN}/S^\ast_{\rm CO}$ ratios are given in Table \ref{tb:ngc604cloudproperties2}.

There is similar GMC's HCN survey in M~31 using IRAM 30-m radiotelescope \citep{2005A&A...429..153B}. 
We found that our estimates of $S_{\rm HCN}/S^\ast_{\rm CO}$ are comparable \footnote{This might infer that the missing fluxes in our interferometric observations are nearly equal between the CO images and the HCN images} to ratios of 0.0075--0.028 that obtained from their spectrum data of M~31 at a spatial scale of $\sim$100\,pc ($\sim$28\arcsec).
The $S_{\rm HCN}/S^\ast_{\rm CO}$ ratios of the two HCN components are also comparable to those of typical galactic GMC \citep[$0.014\,\pm\,0.020$ on average;][]{1997ApJ...478..233H}, which were measured in nearby GMCs at the spatial resolution of 0.5\,pc.

\subsection{89\,GHz continuum}
\label{sec:89}
\begin{figure*}
\begin{center}
\includegraphics[scale=0.5, trim=80 0 0 0, clip]{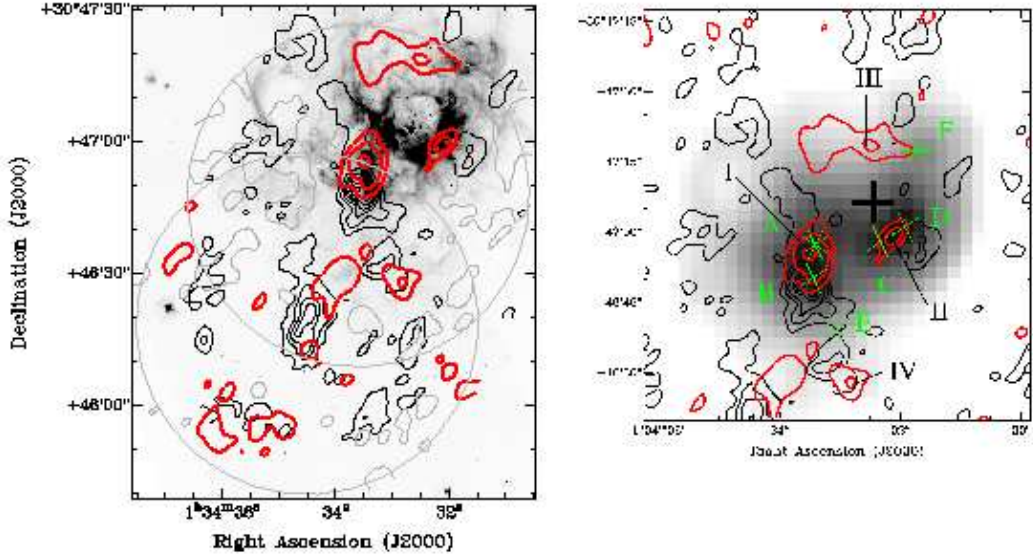}
\caption{({\it left}) The 89\,GHz continuum map (red contour) and CO (black contour) overlaid on the H$\alpha$ (gray scale). The red contour levels for 89\,GHz continuum map are -2, 2, 4, and 6\,$\sigma_{\rm rms}$, where 1\,$\sigma_{\rm rms}=0.6\,{\rm MJy\,sr}^{-1}$, corresponding to 0.6 mJy\,beam$^{-1}$ for the $6 \farcs 9 \times 5 \farcs 5$ resolution data. The negative contours are drawn as gray contour. The black contour levels for CO are 3, 6, 9, 12, 15, 18, and 21\,$\sigma_{\rm rms}$. The gray circles with the diameter of 77\arcsec are two FoVs that we observed in HCN and 89\,GHz continuum emission. The parts of the emission outside the FoVs are masked out. Primary beam correction was applied to this image. ({\it right}) Comparison of 89\,GHz continuum emission (red contour) with 24\,$\mu$m emission (gray scale)  overlaid CO image (black contour). The 89\,GHz source names (I--IV) are labeled. The cross symbol indicates the position of the center of the main cavity in NGC~604. 
The green star symbols represent the peak position of the 8.4\,GHz components (A--F) by \citet{1999ApJ...514..188C}. \label{fg:89gmoment}}
\end{center}
\end{figure*}

\begin{figure*}
\includegraphics*[scale=0.6,trim=0 100 0 0, clip]{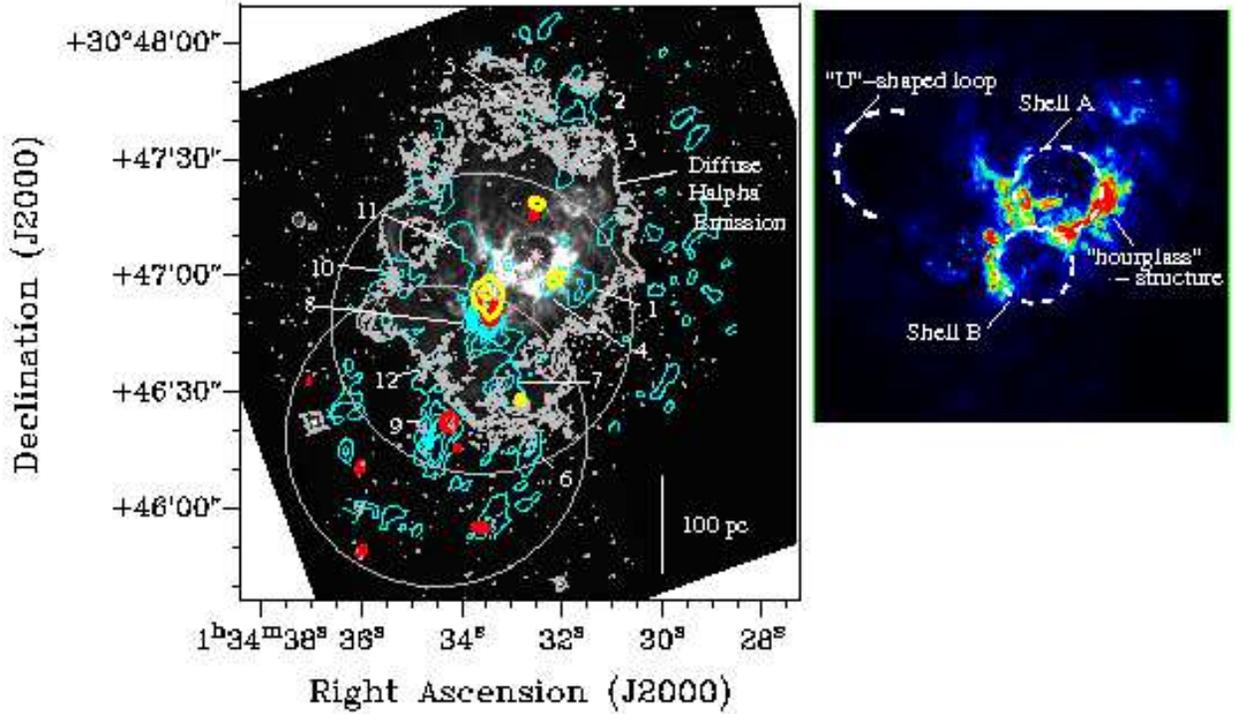}
\caption{({\it left}) Interferometric integrated intensity maps of $^{12}$CO($J$=1--0)
 emission (cyan), HCN emission (red) and 89 GHz
 continuum (yellow) overlaid color image of H$\alpha$ (gray scale) by HST, which are the same images as Figure 1--5. But yellow contour levels are $> 4\,\sigma_{\rm rms}$ and red contour levels are 4, 5\,$\sigma_{\rm rms}$ in order to show clearly the peak positions. The gray contour levels are 0.2\,\% of the peak intensity so as to show the extent of the diffuse H$\alpha$ emission nebula. 
The pink star symbol indicates the position of the center of the main cavity in NGC~604.
The gray circles with the diameter of 77\arcsec are two FoVs that we observed in HCN and 89\,GHz continuum emission. The cloud numbers and linear scale bar are also plotted. ({\it right}) The trimmed image of Figure 1 draws the features of mains shells and annotations discussed in the text. The locations of two shells named A and B are indicated on the map. \label{fg:total}}
\end{figure*}

We detected the 89\,GHz continuum emission in \objectname{NGC~604} for the first time. Four continuum components were detected above 4-$\sigma_{\rm rms}$ noise level in our synthesized image. The two strongest detections ($> 5\,\sigma_{\rm rms}$) in our NMA images are labeled as \ion{Source}{1} and \ion{Source}{2}.
The right panel of Figure \ref{fg:89gmoment} shows the 89\,GHz continuum map overlaid on the CO integrated-intensity map.

\ion{Source}{1} was detected over 6\,$\sigma_{\rm rms}$ at the position ($01^h34^m33\fs5$, $+30\arcdeg46\arcmin55\farcs9$) and appears to be associated with NMA-8 (within 9\arcsec).
\ion{Source}{2} was detected over 5 $\sigma_{\rm rms}$ at the position $(01^{\rm h}34^{\rm m}32\fs1$, $+30\arcdeg46\arcmin59\farcs3)$ which is consistent with the position of the CO emission peak of NMA-4. 
Thus we consider that these two 89\,GHz components can be physically associated with NMA-8 and NMA-4, respectively. 
Moreover, we found that the distribution of these sources, \ion{Source}{1} and \ion{Source}{2}, exhibits a similarity to the 8.4~GHz continuum emission components A--D presented in \citet{1999ApJ...514..188C}; our \ion{Source}{1} corresponds to the position of their A and B components, while our \ion{Source}{2} to the position of their C and D components (see Figure \ref{fg:89gmoment}). The other two components, \ion{Source}{3} and \ion{Source}{4}, are located at ($01^{\rm h}34^{\rm m}32\fs4$, $+30\arcdeg47\arcmin18\farcs8$) and ($01^{\rm h}34^{\rm m}32\fs8$, $+30\arcdeg46\arcmin28\farcs8)$, respectively.
The signal-to-noise ratio of these components is low ($<5\sigma_{\rm rms}$) and \ion{Source}{3} is located at the edge of the FoV. In addition, neither CO clouds nor 8.4~GHz emissions are found to be associated with them. Therefore, we do not consider the \ion{Sources}{3} and \ion{Source}{4} for further discussions.

We measured the flux densities over the region where the emission was seen ($>\,3\,\sigma_{\rm rms}$) and the errors of the total flux densities were estimated taking account the uncertainties in the absolute flux scale and the noise fluctuations. The estimated flux densities of the two components are $3.4\,\pm\,0.9$\,mJy and 9.1\,$\pm\,1.5$\,mJy in the NMA-4 region (\ion{Source}{2}) and in the NMA-8 region (\ion{Source}{1}), respectively. 
The millimeter-wavelength continuum flux provides a constraint on the emission measure of ionized gas, from which we can obtain information about the number of stars present for models of number of UV photons.
This is because the free-free emission from the gas ionized by OB stars contributes largely to millimeter-wavelengths and is optically thin at millimeter-wavelengths.
Other contributions to millimeter-wavelength continuum, non-thermal radiation from supernova remnants (SNRs) and thermal emission from interstellar dust, are expected to be small.
Assuming that the 89\,GHz continuum flux densities are entirely dominated by free-free emission from the ionized gas, we calculated the total production rate of ionizing photons from the observed 89\,GHz continuum flux densities.

In the calculation, we followed equation (2) in \citet{1991ApJ...366L...5S}, which provides the relationship of the free-free emission flux density at 110\,GHz to the total production rate of ionizing photons for optically thin plasma. 
Using the thermal (free-free) spectral index of $-0.1$ in M~33 \citep{1988A&A...205...29B}, the relationship of the free-free emission flux density at 89\,GHz ($S_{\rm 89GHz}$) and the total production rate of ionizing photons (Q$^*$s$^{-1}$) is given by:
\begin{equation}
S_{\rm 89GHz}=8.58 \times 10^{-2} \frac{Q^*/10^{49}{\rm s}^{-1}}{d^2_{\rm Mpc}} {\rm mJy},
\end{equation}
where $d_{\rm Mpc}$ is the distance to M~33. The H\,{\sc ii} regions are assumed to be ionization-bounded at a temperature of 10$^4$ K. The total production rate of ionizing photons are $(28.0\,\pm\,7.3)\times10^{49}$\,s$^{-1}$ in the NMA-4 region (\ion{Source}{2}) and $(74.8\,\pm\,12.3)\times10^{49}$\,s$^{-1}$ in the NMA-8 region (\ion{Source}{1}).
The value for \ion{Source}{1} is consistent with the total value expected from the 8.4~GHz fluxes of components A and B \citep[68$\times10^{49}$\,s$^{-1}$;][]{1999ApJ...514..188C}. As for \ion{Source}{2} it is not consistent with C and D components (64$\times10^{49}$\,s$^{-1}$). It is possible that more flux from large scale structures in NMA-4 are resolved out than those in NMA-8, because missing flux in interferometric observations depends on the sizes of the emission sources.
Assuming that the missing flux at 89\,GHz continuum is the same as that of the CO images, the ionizing photon production rate for the NMA-4 region is consistent with that from the 8.4~GHz fluxes of components C and D.

\subsection{Overall features and Comparison with H$\alpha$}
\label{sec:hst}
In this section we compare the distributions of CO, HCN, and 89\,GHz continuum emissions with the morphology of the H$\alpha$ shells in NGC~604.
The composite image of CO and HCN line emissions, and 89\,GHz continuum emission overlaid onto the H$\alpha$ image obtained by the {\it Hubble Space Telescope} (HST) is shown in Figure \ref{fg:total}.

Three out of ten identified clouds (and NMA-11) are associated with the bright shells, i.e., Shell~A and Shell~B (see Figure \ref{fg:total}). 
The bright western part of Shell~A shows an hourglass-like structure, which appears to be fitted by the distribution of $^{12}$CO emission of NMA-4.
The velocity structures of these H$\alpha$ shells show true expansion \citep{1996AJ....112..146Y}, possibly powered by stellar winds from the central OB star clusters and/or SNRs (Section 1). 
The line-of-sight velocity of NMA-4 ($-247$\,km\,s$^{-1}$), which corresponds to that of the H$\alpha$ shell at the same position (H$\alpha$ emission reported in \citet{2000ApJ...541..720T}), corresponding to Shell~A. Therefore, it is considered that NMA-4 is a cloud component associated with the expanding H$\alpha$ shell, Shell~A. The cloud NMA-11, the marginally detected cloud, is also located on the eastern part of Shell~A.

The cloud NMA-8 itself has an arc-like structure and is distributed to surround the cavity, nicely tracing the bright rim of Shell~B. 
The 89\,GHz continuum and HCN emission, which are considered to be associated with NMA-8, were detected in the inner wall of Shell~B, i.e. to the direction of the central stellar cluster. Moreover, there is an offset between peak positions of three probes: CO, HCN, and 89\,GHz continuum, which are aligned to the direction of central stellar clusters (from south to north).
The line-of-sight velocity of NMA-8 ($-243$\,km\,s$^{-1}$) corresponds to that of the H$\alpha$ shell at this position \citep{2000ApJ...541..720T}, corresponding to Shell~B. Therefore, it is considered that NMA-8 is a cloud component associated with the expanding shell B.

Looking toward the halo structure of the H$\alpha$ nebula, the distribution of $^{12}$CO emission of NMA-2, NMA-6, NMA-9, and NMA-12 do not overlap with the area seen in H$\alpha$ emission but extend along the edge of diffuse H$\alpha$ emission. Although it is difficult to address whether or not they are physically associated with the diffuse H$\alpha$ emission due to lack of the spectral data of H$\alpha$, we would like to mention the interesting feature of the dense-gas-forming region: the HCN component detected around the NMA-9 cloud, showing offset between CO peak and HCN peak, is located in the northern part of NMA-9; i.e., to the direction of the H$\alpha$ nebula. 
As for the rest of the clouds, they are seen on the diffuse H$\alpha$ emission. 
The cloud NMA-10 is located in the southern part of the ``U''-shaped H$\alpha$ loop, which marginally shows expansion signatures \citep{2000ApJ...541..720T}.

Consequently, we found the distribution of CO, HCN, and 89\,GHz continuum emissions which we detected in our observations are well correlated with the expanding H$\alpha$ shells in NGC~604, in both morphology and kinematics. 
In contrast with the concentration of molecular gas on the rim of shells, no molecular clouds are observed inside Shells A and B.

\section{Discussion}
\label{sec:4}
\subsection{Comparison with 24\,$\mu$m and H$\alpha$ emission}
In this section, we derive the star formation efficiencies (SFEs) from the available {\it Spitzer} Archival 24\,$\mu$m data and H$\alpha$ data in order to discuss the star formation activities over all five FoVs, because a combination of H$\alpha$ and 24\,$\mu$m luminosities is a better tracer of the total star formation rate \citep[SFR,][]{2007ApJ...671..333K}. 
We then compared the SFEs with those in inferred from our 89\,GHz data.

The 24\,$\mu$m emission is often used as an indicator of the current star formation because it is radiated from dust grains heated by newly formed massive stars \citep[e.g.,][]{2009ApJ...699.1125R}. The H$\alpha$ emission is also directly linked to star formation, however, it is only sensitive to the ionizing photons unabsorbed by dust. 
There is empirically known to be a linear combination of the 24 $\mu$m emission, tracing the obscured star formation, and the observed H$\alpha$ luminosity, tracing the un-absorbed star formation, which correlates better than other SFR tracers with the extinction-corrected H$\alpha$ luminosity \citep{2007ApJ...666..870C}.

We retrieved the 24\,$\mu$m images of M~33 obtained with MIPS \citep{2004ApJS..154...25R} from the {\it Spitzer} Space Telescope \citep{2004ApJS..154....1W} data archive. The Basic Calibrated Data (BCD) was created using the {\it Spitzer} Science Center (SSC) pipeline. A background matching was applied between overlapping fields of view and background gradient was removed. The FWHM of the point spread function in the image is $5\farcs7$ and the grid size of the final image is $2\farcs5$ per pixel.
Our 89\,GHz continuum image was overlaid on the resultant 24\,$\mu$m image as shown in Figure \ref{fg:89gmoment}. Both \ion{Source}{1} and \ion{Source}{2} in our 89\,GHz continuum image correspond to peaks in the 24\,$\mu$m map.
This is one of the evidences that the 24\,$\mu$m and the 89\,GHz continuum emission both trace sites of star formation embedded in dust clouds.

The reduced data of H$\alpha$ emission in M~33 were kindly provided by R. Walterbos. It was observed with the 0.6-m Burrell-Schmidt telescope at Kitt Peak National Observatory. The dimensions of the CCD are 2048$\times$2048 with pixels of 2$\farcs$03 and a total field of view of about $70\arcmin\times70\arcmin$, and then trimmed with the same size of our CO image in NGC~604. The sensitivity is $0.8\times 10^{−17}\,{\rm erg\,s^{-1}\,cm^{-2}\,arcsec^{-2}}$ \citep{2001ApJ...559..878H}. More details of the observations and reduction process are described in \citet{2000ApJ...541..597H}. 

\subsubsection{SFE from 89\,GHz continuum emission}
The total stellar mass can be compared with the mass of the molecular clouds to measure SFE. 
The SFE$_{\rm 89GHz}$ is defined as follows:
\begin{equation}
{\rm SFE_{\rm 89GHz}}=\frac{M_{\rm star}}{M_{\rm star}+M^*_{\rm mol}},
\end{equation}
where $M_{\rm star}=\int_{M_{\rm min}}^{M_{\rm max}}N(M) M\,dM$ is total stellar mass between $M_{\rm min}$ and $M_{\rm max}$, and $N(M)$ represents stellar mass function. 
We defined the missing-flux-corrected molecular cloud mass as $M^\ast_{\rm mol}$ using the missing flux derived in Section \ref{sec:2}.
We analyzed total number of stars from the total ionizing photon rates (Section \ref{sec:89}) using the ionizing rates of each O-type stars \citep[\ion{O9.5}{5} -- \ion{O3}{5}][]{2005A&A...436.1049M}, because these stellar spectral types are considered to be largely responsible for the ionizing photons of interstellar matter.
We assumed that the H\,{\sc ii} region NGC~604 have Salpeter IMFs
\citep[2.35;][]{2000MNRAS.317...64G} with mass limits of 0.1--58\,M$_{\sun}$ \citep[corresponds up to spectral type \ion{O3}{5};][]{2005A&A...436.1049M}.
The number of O-type stars (\ion{O9.5}{5} -- \ion{O3}{5}) in the NMA-4 and NMA-8 region is 185$\pm$31 and 70\,$\pm$\,18 stars, respectively. The SFE$_{\rm 89\,GHz}$ was estimated to be $12.4\,\pm\,6.6\,\%$ for NMA-4 and $5.1\,\pm\,2.3\,\%$ for NMA-8. As for other clouds which were not detected in continuum emission, upper limits of SFE$_{\rm 89\,GHz}$ were calculated and they are given in Table \ref{tb:ngc604cloudproperties2}.

\subsubsection{SFE from 24\,$\mu$m emission and H$\alpha$ emission}
\label{sec:4.1.2}
The H$\alpha$ and 24\,$\mu$m images were convolved into a common 5\farcs7 resolution and regridded to 2\farcs5 per pixel.
We measured the 24\,$\mu$m and H$\alpha$ luminosity over the same spatial regions of all the CO clouds we identified.  
We measured the integrated flux over the extent of H$\alpha$ nebula and compared with previously reported fluxes from the literature \citep{1984ApJ...287..116K}. 
The H$\alpha$ fluxes in \citet{2009ApJ...699.1125R} are computed from the observed H$\alpha+$[\ion{N}{2}] fluxes using the spectroscopically determined [\ion{N}{2}]/H$\alpha$ ratios \citep{2002MNRAS.329..481B}. 
The extinction–corrected H$\alpha$ luminosity is then expressed
as a linear combination of these two luminosities: $L_{\rm corr}({\rm H}\alpha)=L_{\rm obs}({\rm H}\alpha)+a\times L({\rm 24\,\mu m})$, where $a=(0.031\pm0.006)$ \citep{2007ApJ...666..870C}.
Ionizing photon rates are derived from extinction-corrected H$\alpha$ luminosities ${\rm L(H\alpha_{corr})}$ using 
\begin{eqnarray}
Q^{\ast} (s^{-1}) = 7.32 \times 10^{11} L_{\rm corr}({\rm H}\alpha)\,({\rm erg\,s^{-1})},
\end{eqnarray}
which assumes an ionization bounded case B nebula \citep[][and reference therein]{1996AJ....111.1252M}.
The star formation efficiency from 24\,$\mu$m and H$\alpha$ (SFE$_{\rm H\alpha,24\mu m}$) was calculated using the same procedures described in previous subsection 4.1.1.  The SFE$_{\rm H\alpha,24\mu m}$ for each cloud is given in Table \ref{tb:ngc604cloudproperties2}. The NMA-4 has the highest SFE of all clouds and there is the same tendency in both SFEs. The derived SFE$_{\rm H\alpha,24\mu m}$ of 0.2 -- 11.5\,\% is good agreement with that from 89\,GHz continuum ($<\,1.2$ -- 12.4\,\%) within errors.

The total amount of ionizing photon rate ($Q^{\ast}$) in the NGC~604 region as a whole (within a radius of 400\,pc) is 413 photons s$^{-1}$. The summation of the $Q^{\ast}$ over identified clouds is 188 $\times 10^{49}$ photons s$^{-1}$, which occupies about 46\,\% of the total $Q^{\ast}$ of a whole region. On the other hand, the $Q^{\ast}$ from the central stellar clusters ($\sim$ 200 O~stars) is estimated to be 145 $\times 10^{49}$ photons s$^{-1}$, assuming that the central stellar clusters have the Salpeter IMF (2.35) and \citet{2005A&A...436.1049M}'s ionizing photon rates for each stellar spectral type, which contributes about 35\,\% to the total $Q^{\ast}$ of a whole region. Thus this implies that such a large ionized region has been created not only because of the central stellar cluster but also the embedded stellar clusters in the clouds.

\subsubsection{Comparison to other H\,{\sc ii} regions}
In comparison, we derive SFEs of three well-known H{\sc ii} regions (Orion A, W49A, and 30 Doradus) in the Local Group. 
The first two H{\sc ii} regions contain similar molecular cloud mass (i.e., mass of 1--7$\times10^5$\,M$_{\sun}$) to our identified CO clouds in NGC604. Orion A is the nearest and most well-studied region of star formation in our Galaxy, and is known to be excited by a few late O-type star or early B-type star \citep[e.g.,][]{1988AJ.....95..516C}. 
W49A is one of the most massive and luminous H\,{\sc ii} regions in our Galaxy and contains 30 individual ultracompact H\,{\sc ii} regions, corresponding to about that number of OB stars \citep{2002ApJ...564..827C}. 
The most luminous in the Large Magellanic Cloud (LMC), 30 Doradus giant H{\sc ii} region is introduced to compare with the secondary luminous H{\sc ii} region, NGC~604. Very luminous NGC~604-class H\,{\sc ii} regions is absent in our Galaxy, and hence it is frequently compared to the NGC~604, although 30 Doradus has about 3--4 times as total ionizing flux as NGC~604. 
The estimated SFEs with the same assumption above are tabulated in Table \ref{tb:SFE}. 
The $Q^{\ast}$ of Orion A was determined from the stellar types in \citet{1988AJ.....95..516C} and the SFE of Orion A is 2.2\,\%. 
The $Q^{\ast}$ of W49A was calculated from the measurement of combining the 25$\mu$m luminosity \citep{1990MNRAS.244..458W} and H$\alpha$ emission \citep{1984ApJ...287..116K}, and this can be a value of 182$\times10^{49}$\,s$^{-1}$. 
The $Q^{\ast}$ of 30~Doraus is derived only from H$\alpha$ emission \citet{1984ApJ...287..116K}, and thus the SFE of 30~Doradus can be more than 52.8\%.
The SFE of NMA-4 is much larger than Orion A and comparable to those of the galactic H{\sc ii} region W49A, while the SFE of NMA-8 is comparable to or slightly less than W49A.
The SFE of each decomposed cloud in supergiant H\,{\sc ii} region NGC~604 is quite similar to that of galactic H\,{\sc ii} region. Both of NMA-4 and NMA-8 have about three to ten times lower SFEs than 30~Doradus, although they are needed to be compared on the same scale.

The averaged efficiency of star formation out of giant molecular associations \citep[GMAs; e.g.,][]{1990ApJ...349L..43R} in other extragalactic GHRs tends to be about 1--10\,\% out of the parental molecular cloud mass \citep[10$^6$--10$^7$\,M$\sun$,][]{1996AJ....111.1252M}.
In order to compare SFEs of them with that of NGC~604 on the same scale, the total (missing-flux-corrected) mass in NGC~604 region is estimated to be 51 $\times 10^5$\,M$_{\sun}$, and the extinction-corrected H$\alpha$ luminosities $L_{\rm corr}({\rm H\alpha})$ is to be $4.3\times10^{39}$\,s$^{-1}$. The derived SFE over NGC~604 region is 7\,\%. This is good agreement of other extragalactic GHRs. On the other hand, the SFE of the most luminous H\,{\sc ii} region 30 Doradus is extremely higher than NGC~604 or any other GHRs, which suggests its half of parental molecular clouds have been converted into stars and also dozens of O-stars would have dissipated a whole GMC. NGC~604 may have potential to continue star formation activity, which leads to a 30~Doradus-scale GHR, since NGC604 has still contains a considerable reservoir of molecular gas to be used for star formation.
 
\subsection{The Different Properties and Evolution of GMCs}
\label{sec:4.2}
As we mentioned in the former section, we found that the identified clouds have different SFEs (see Table \ref{tb:ngc604cloudproperties2}). 
Especially, NMA-4, NMA-8, and NMA-9 have conspicuous differences in their SFE$_{\rm H\alpha, 24\mu m}$.
In this section, we focus on these clouds and discuss the relationship between their different properties and their evolutionary stages.

\subsubsection{NMA-4 and NMA-8} 
NMA-4 is located near the central stellar clusters (the projected distance from the clusters is $\sim$32\,pc) and also along the bright rim of Shell~A (western part of Shell~A; see also Figure \ref{fg:total}).
The cavity surrounded by Shell~A is known to be filled with hot coronal gas that emits X-rays and is transparent to the ionizing UV \citep{2004AJ....128.1196M,2008ApJ...685..919T}. 
The 8\,$\mu$m emission, which is often used as tracer of photo-dissociation region, is mostly distributed around the UV emission \citep[see Figure 2 in][]{2009ApJ...699.1125R}, where the clouds NMA-8 and NMA-4 are located.
The NMA-4 and a part of NMA-8 are considered to be directly exposed to the UV radiation from young massive stars. 

In comparing the physical properties of NMA-4 with NMA-8, we found significant enough to distinguish between their evolutionary stages. 
The clouds NMA-4 and NMA-8 have the highest SFE and the second-highest SFE, respectively, and nevertheless, they are different by a factor of $\sim 3$. Its significant difference in SFEs is mainly due to their respective masses of molecular gas; NMA-4 has the cloud mass of $\sim$0.8${\times}$10$^5$ M$_{\sun}$ and size of $\sim$10\,pc, while NMA-8 has the cloud mass of $\sim$7.4${\times}$10$^5$\,M$_{\sun}$ and size of $\sim$29\,pc. These results might suggest that most of the cloud mass in NMA-4 has been more affected with star formation activities, i.e., conversion into stars, dissipation by the stellar winds, or disruption by intensive photo-dissociation, than in NMA-8. The cloud NMA-8 is one of the largest molecular clouds and shows the obvious sign of dense gas formation (Section \ref{sec:3.2}). 
Since NMA-8 still contains a considerable reservoir of molecular gas and the dense gas that has not yet been dissipated by UV radiation, it is likely to undergo star formation activities. 
The cloud NMA-8 can be at a stage of ongoing massive star formation as well as dense gas formation.

\subsubsection{NMA-9}
This is the second largest cloud (size of $\sim28$\,pc and mass of $\sim6.4\times10^{5}\,{\rm M_{\sun}}$) with the lowest SFE. Its low SFE$_{\rm 24\mu m}$ suggests a lack of recent massive star formation. This cloud includes dense molecular gas, which is suggested by the detection of HCN line emission.  Therefore, NMA-9 might be just forming dense molecular gas in the GMC but not be led to form stars yet.

\subsubsection{The other clouds}
The other clouds show much lower values in SFE$_{24\mu m}$ than NMA-8 and NMA-4.
We are not sure whether they include dense gas or not because HCN emissions were not detected over 3\,$\sigma_{\rm rms}$ in some clouds (NMA-1, NMA-6, NMA-7, and NMA-10).  The $S_{\rm HCN}/S^\ast_{\rm CO}$ ratio upper limits allow for the possibility that CO clouds have dense gas fraction similar to typical GMCs \citep{1997ApJ...478..233H,2005A&A...429..153B}. Much deeper HCN observations are necessary to distinguish different properties from the dense gas fraction.
Even though as for other clouds (NMA-2, NMA-3, and NMA-5), they have not been observed in HCN emission, we briefly describe the possibility of dense gas formation in the third largest cloud, NMA-2, since CO(3--2) emission was detected \citep{2007ApJ...664L..27T}.
The CO(3--2) is often used as a tracer of warmer and/or denser molecular gas ($n_{\rm H_2}\sim10^4$\,cm$^{-3}$, $T_{\rm crit}=33\,{\rm K}$) than CO(1--0). In fact, CO(3--2) is a better tracer of density than temperature in the region where the CO(3--2)/CO(1--0) ratio is less than 0.7 when assuming the kinetic temperature is low \citep[$\sim$ 20\,K;][]{2007PASJ...59...43M}.
Its CO(3--2)/CO(1--0) ratio ($\sim 0.5$) has a value similar to that of the second largest cloud NMA-9 \citep{2007ApJ...664L..27T}.
Since NMA-9 shows no sign of star formation but dense gas formation in our results, the detection in CO(3--2) emission on this cloud could be due to the dense gas formation rather than heating by embedded young stars.
Considering that NMA-2 has low SFE$_{\rm H\alpha,24\mu m}$ and shows little sign of star formation, NMA-2 is likely to have a similar status regarding dense gas and star formation, i.e., the clouds include a significant fraction of dense gas but have low star formation activities.

In summary, NMA-9, NMA-8, and NMA-4, seem to have significant different properties in terms of dense gas formation and massive star formation. This variety can be explained as differences in their evolutionary stages, from dense gas formation to star formation.
Cloud NMA-9 could be at the stage of ``dense gas formation with no massive star formation''. Cloud NMA-8 could be at more evolved stage of ``dense gas and ongoing massive star formation'' than NMA-9. Cloud NMA-4 could be at the most evolved stage of ``remnant'' cloud where most of the molecular gas has already been converted into stars and been dispersed by newly born massive stars. 

\subsection{Sequential Star Formation in NGC~604}
\label{sec:4.3}
As we mentioned in the previous section, we found a distinct difference in the physical states of star formation among the GMCs in the NGC~604 region and explained it as a difference in their evolutionary stages.
Furthermore, these GMCs are well correlated with the spatial structure, and some of them are also with kinematics, of the H$\alpha$ nebula including the expanding shells (Section \ref{sec:hst}). 
The physical properties and the evolution of GMCs are straightforwardly expected to be affected by the H\,{\sc ii} region excited by the stellar activities of OB clusters. 
The Shell~B might have been formed by the blowout through the molecular tunnel over Shell~A \citep{2000ApJ...541..720T} and is estimated to be younger than Shell~A \citep{1996AJ....112..146Y}, which is good agreement with that NMA-8 is at a younger stage than NMA-4 because NMA-4 has already dissipated parent molecular cloud as mentioned in Section 4.2.
These results are explained with the scenario of sequential star formation in NGC~604 that was previously proposed \citep{2007ApJ...664L..27T}.

\begin{figure*}
\begin{center}
\includegraphics[scale=0.45]{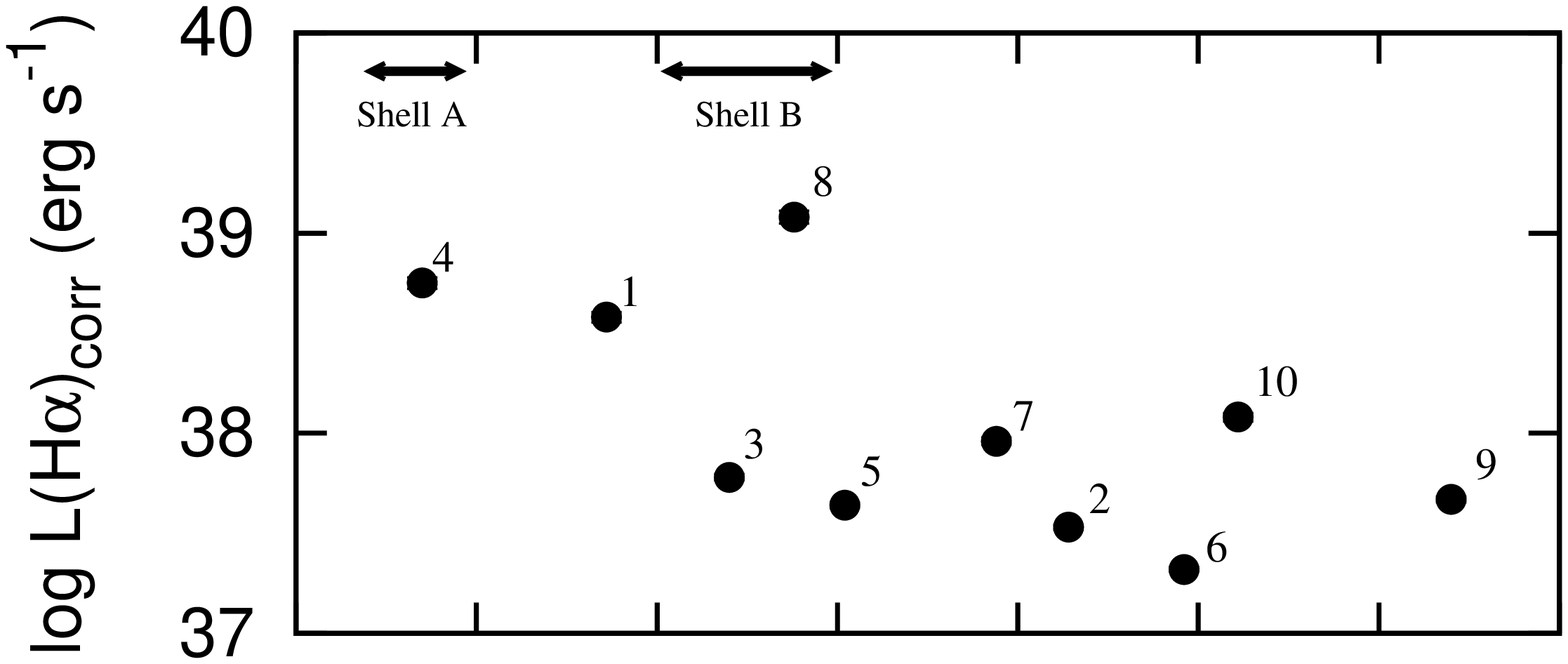}\\
\includegraphics[scale=0.45]{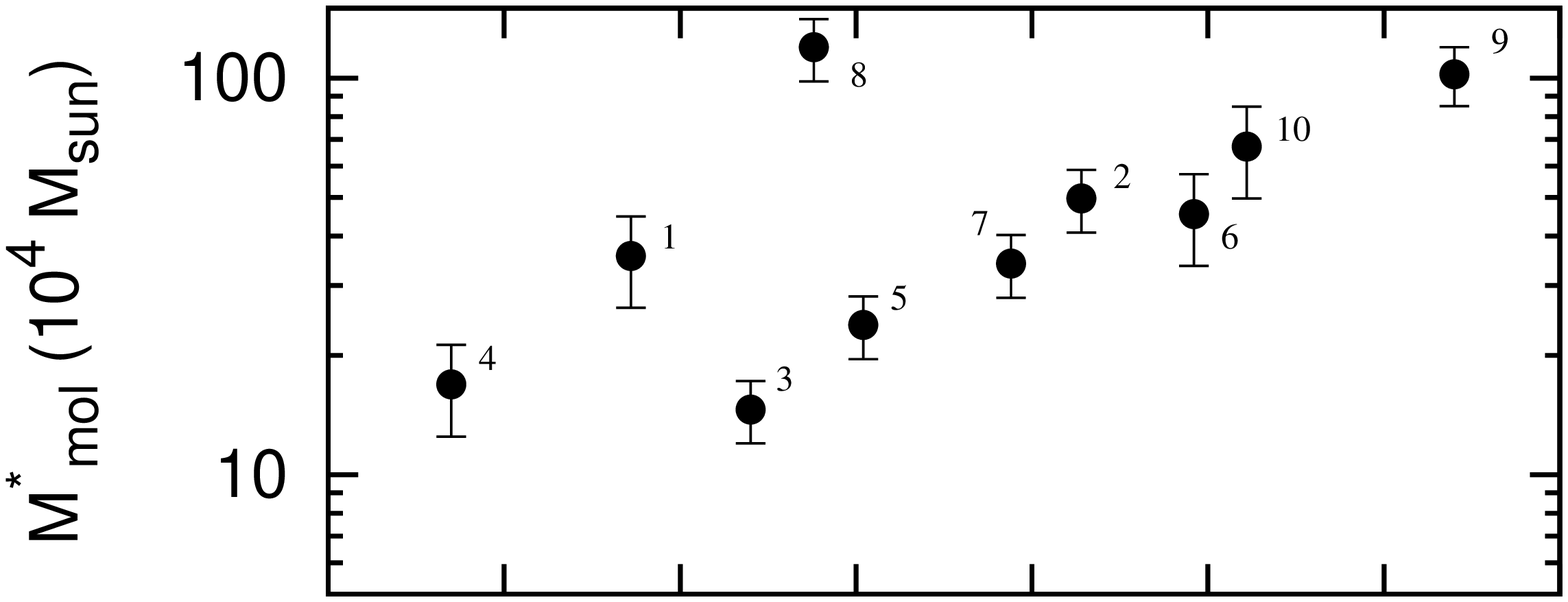}\\
\includegraphics[scale=0.6]{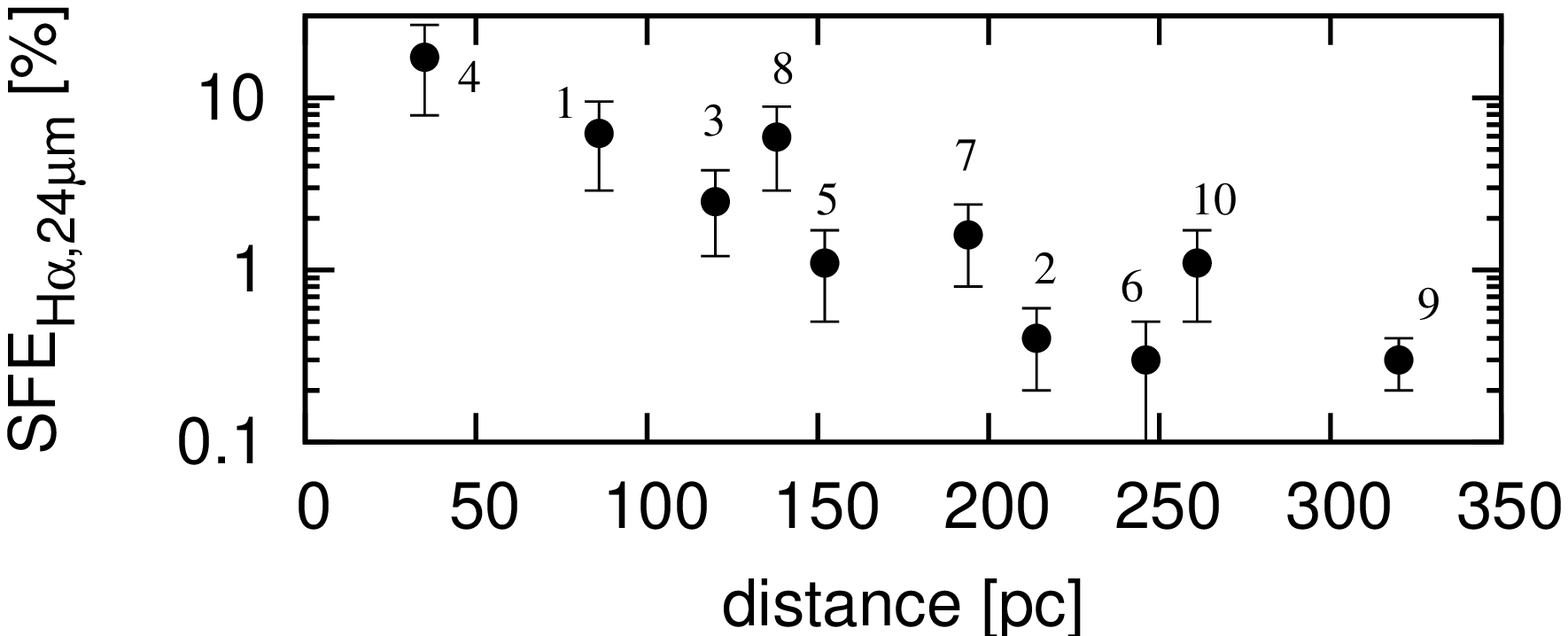}\\
\caption{L(H$\alpha$)$_{\rm corr}$ ({\it top}), M$^\ast_{\rm mol}$ ({\it middle}), and SFE$_{\rm H\alpha,24\mu m}$ ({\it top}) distribution as a function of projected distance from the center of the main cavity in NGC~604. Its coordinate of ($01^{\rm h}34^{\rm m}32\fs5$, $+30\arcdeg47\arcmin0\farcs5$) is indicated as a pink star symbol in Figure \ref{fg:total}. The cloud numbers  are also plotted. In the top panel, the location of two H$\alpha$ shells are indicated for reference. Note that common errors are excluded in order to compare each clouds relatively. \label{fg:SFE2}}
\end{center}
\end{figure*}

Figures \ref{fg:SFE2} shows the radial distribution of $L$(H$\alpha$)$_{\rm corr}$, M$^{\ast}_{\rm mol}$, and SFE$_{\rm H\alpha,24\mu m}$ for each cloud as a function of projected distance from the center of the main cavity in NGC~604 ($01^{\rm h}34^{\rm m}32\fs5$, $+30\arcdeg47\arcmin0\farcs5$), where the central clusters are crowded most. The gradual decrease of $L$(H$\alpha$)$_{\rm corr}$ as the radial distance increases is found in Figure \ref{fg:SFE2} (top), except the NMA-8 region (Shell~B region). 
On the other hand, we found the gradual increase of M$^{\ast}_{\rm mol}$, which implies that the star formation are active enough to consume and/or to dissipate parental clouds. We found a clear radial dependence in SFE$_{\rm H\alpha,24\mu m}$; the closer to the central clusters the GMCs are located, the higher SFE they have. 
We should note that this is because of the combination of the radial dependency of both $L$(H$\alpha$)$_{\rm corr}$ and M$^{\ast}_{\rm mol}$ rather than only of $L$(H$\alpha$)$_{\rm corr}$ distribution. Besides, the total flux of $L$(H$\alpha$)$_{\rm corr}$ is considered to be contributed from both the central clusters and the young stars that might be embedded in the clouds (Section \ref{sec:4.1.2}).  
As for NMA-2, NMA-6, and NMA-9 we don't know whether they are physically associated with H$\alpha$ emission or not. In the case that they are observed at associations by chance, the values of SFE that we derived give just the upper limits. Nevertheless, the SFE on the bottom panel in Figure \ref{fg:SFE2} still shows tendency that the SFE decrease as the distance from the central clusters increase.

\begin{figure*}
\begin{center}
\includegraphics[scale=0.6]{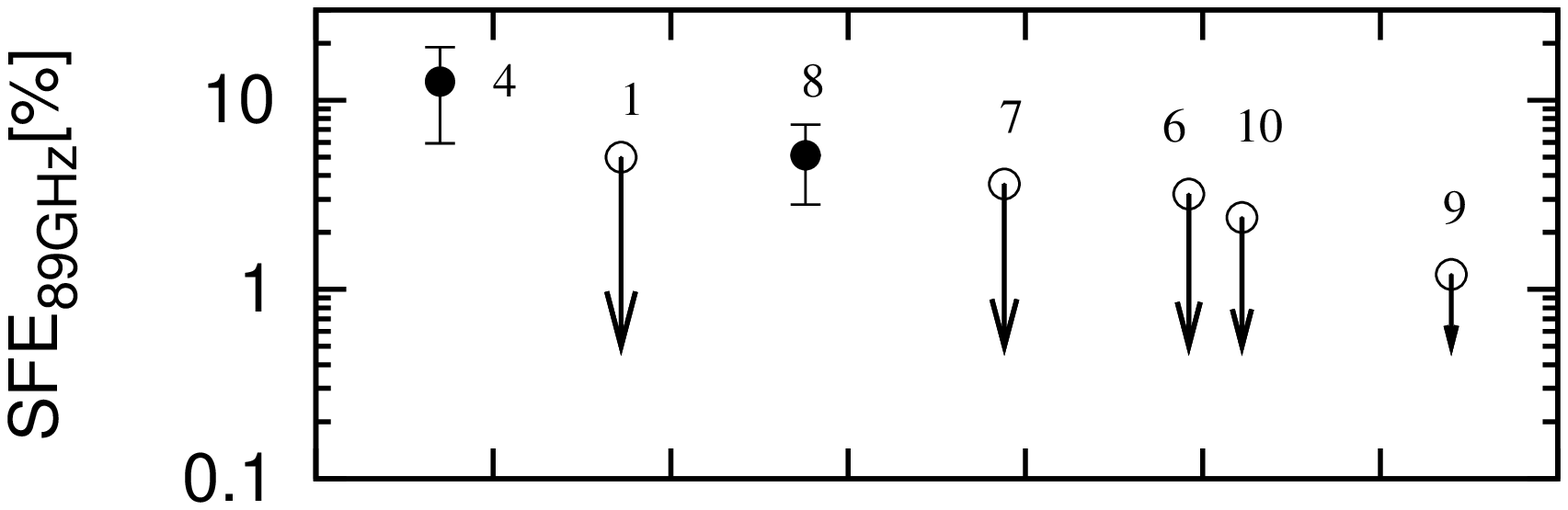}\\
\includegraphics[scale=0.6]{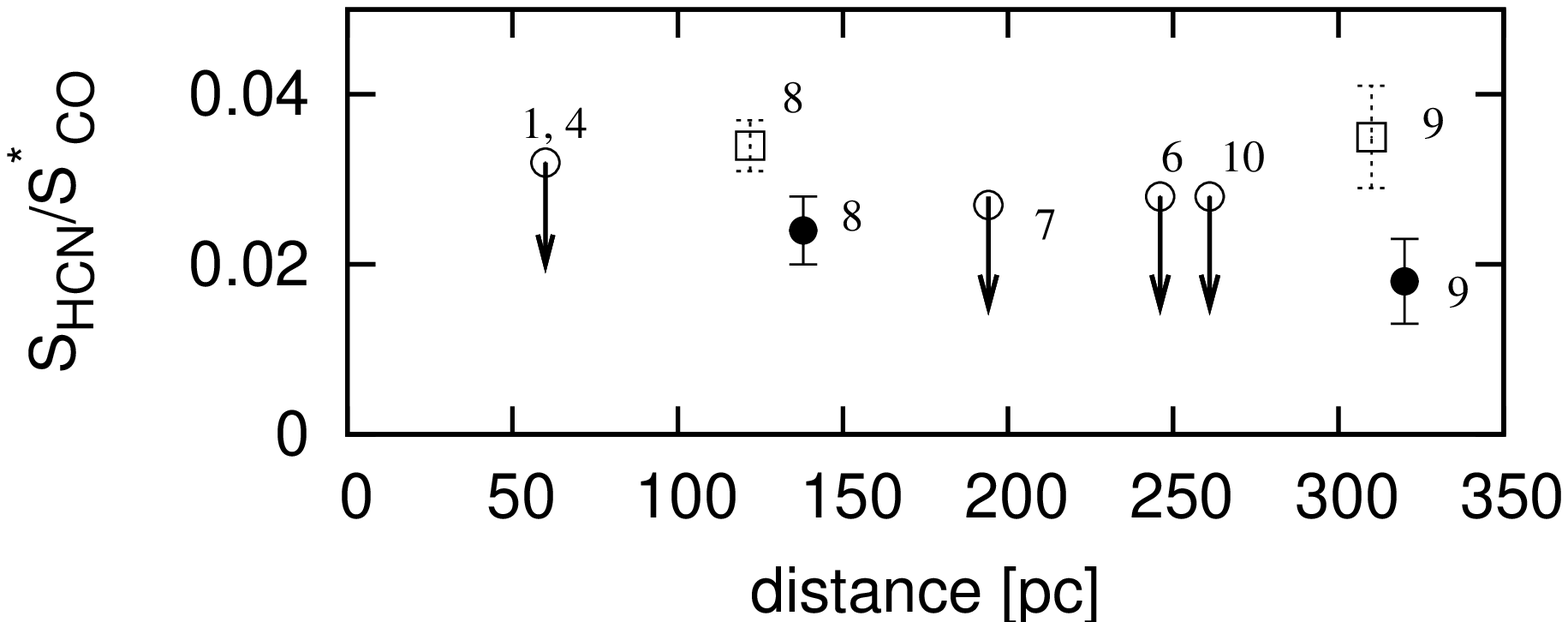}\\
\caption{SFE$_{\rm 89\,GHz}$ ({\it top}), and $S_{\rm HCN}/S^{\ast}_{\rm CO}$ ({\it bottom}) distribution as a function of projected distance from the center of the main cavity in NGC~604. Its coordinate of ($01^{\rm h}34^{\rm m}32\fs5$, $+30\arcdeg47\arcmin0\farcs5$) is indicated as a pink star symbol in Figure \ref{fg:total}. The filled circles represent the detected clouds, while the open circles show the upper limit for the clouds where were not detected in HCN and 89\,GHz continuum emission. The open squares represent the line ratio that measured in the region where HCN emissions were detected. The distance of these two plots were measured at the peak of HCN emission. The cloud numbers are also plotted. Note that common errors are excluded in order to compare each clouds relatively. \label{fg:SFE}}
\end{center}
\end{figure*}

Figure \ref{fg:SFE} also shows the radial distribution of $S_{\rm HCN}/S^\ast_{\rm CO}$ and SFE$_{\rm 89GHz}$ as a function of projected distance from the center of the main cavity in NGC~604.
We found a radial dependence in SFE$_{\rm 89GHz}$, which are in good agreement with the behavior of SFE$_{\rm H\alpha,24\mu m}$ (the bottom panel of Figure \ref{fg:SFE2}), even though these values were derived from different wavelength data. This implies that 89~GHz continuum can be another tracer of massive-star-forming region. 
 On the other hand, the radial dependency of the dense gas fraction $S_{\rm HCN}/S^\ast_{\rm CO}$ is not clear from our data where we have estimated upper limits. 
The dense gas fraction $S_{\rm HCN}/S^\ast_{\rm CO}$ in Figure \ref{fg:SFE} (bottom) is averaged over the extent of the CO clouds (denoted as circles).
However, our high resolution images of CO and HCN emissions in Figure \ref{fg:hcnmoment} notify us that the line ratio of $S_{\rm HCN}/S^\ast_{\rm CO}$ shows a gradient within a single cloud because two HCN components were detected at the northern part of the CO clouds NMA-8 and NMA-9 rather than the peak of the CO emission (see Section \ref{sec:3.2}).
The line ratios of $S_{\rm HCN}/S^\ast_{\rm CO}$ (and its locations) that were measured over the region where HCN emissions were detected are represented as an open square in the bottom panel of Figure \ref{fg:SFE}. 
Their values are higher than the averaged value which are represented as black circles in Figure \ref{fg:SFE} and their locations are close to the direction of the central clusters as compared with the line ratios within a single cloud. Thus from this figure it is found that the dense gas formation has been occurred at the near side of a cloud toward the central clusters.

In the light of our high-resolution observational results, we reconsider the ``sequential star formation" scenario in NGC~604 here in some more detail than the previously proposed scenario \citep{2007ApJ...664L..27T}. First, central clusters were formed, whose powerful stellar wind made the main cavity with Shell~A; this is referred to as ``first-generation star formation'' in \citet{2007ApJ...664L..27T}.
Inside Shell~A, molecular gas around the central part of the stellar clusters is almost cleared out \citep[Section 3.4;][]{2004AJ....128.1196M}. The ``remnant'' cloud, i.e., NMA-4, is on the rim of Shell~A. It is suggested that NMA-4 is a cloud swept up by the stellar wind from the central stellar cluster.
The second-generation stars are formed in the swept-up cloud through dense gas formation and such new stars ionize the surfaces of the parental clouds.

The cloud with ``dense gas and ongoing massive star formation'', i.e., NMA-8, is located on the edge of Shell~B, while molecular gas inside Shell~B is almost cleared out. Similar to the correlation between Shell~A and NMA-4, NMA-8 could be formed with the ambient gas swept-up on Shell~B and compressed by stellar winds. In 
such a compressed dense gas on Shell~B, dense gas, and then stars are newly formed as is evidenced by that dense gas and star forming regions are located at the inner wall of NMA-8 toward the central stellar cluster. Besides, the star forming region in the NMA-8 is much closer to the central stellar cluster than dense gas forming region, and hence, central cluster, massive star forming region, dense gas forming region, and molecular cloud are aligned in this order.
The newly born stars in NMA-8 associated with Shell~B are the latter generation stars of sequential star formation which are induced by the previous generation stars; they are ``second-generation stars'' as referred to \citet{2007ApJ...664L..27T}, or latter generation stars.

On the other hand, the ``dense gas forming'' cloud, i.e., NMA-9, is located on the edge of the extended diffuse H$\alpha$ emission up to a few hundreds of parsecs south of the central cluster. The dense gas forming region which is associated with NMA-9 cloud, are located to the direction of central stellar cluster rather than the emission peak of NMA-9 cloud. 
The expansion of the H\,{\sc ii} region into surrounding medium can trigger star formation through dense gas formation \citep[e.g.,][]{1977ApJ...214..725E}. 
The NMA-2 is located in the north of the diffuse H$\alpha$ nebula, while NMA-9 is in the south of nebula (Section 4.2). 
Their projected distances from the central stellar cluster are almost the same.
If we suppose that the propagation initiated from the central stellar cluster is isotropic, dense gas is expected to be formed in NMA-2 as well as in NMA-9.
The NMA-9 might be compressed by the expansion of the H\,{\sc ii} region.

In summary, we explain the following scenario of sequential star formation in NGC~604. 
The first-generation stars (the central clusters) are formed in a GMC. The central clusters swept up the matter to form Shell~A, where dense gas was formed by compression of the cloud material. Then second-generation stars form in the dense gas, which have already been dissipated in the parental clouds (the ``remnant'' cloud; NMA-4). Shell~B is formed by blowout through the molecular tunnel, where another star formation occurs on the ``on-going dense gas and star formation'' cloud (NMA-8). At the far end beyond these shells dense gas is formed in the quiescent molecular clouds (NMA-9) by the effect of the expansion of the H\,{\sc ii} region.

\section{Summary}
\label{sec:5}
We presented the highest sensitivity and the highest resolution $^{12}$CO images of
 the giant H\,{\sc ii} region NGC~604 in the nearby galaxy M~33. 
Moreover, we also obtained the HCN and 89\,GHz continuum images. This is the first detection of HCN and 89\,GHz continuum emission toward NGC~604 reported in the literature. 

\begin{enumerate}
\item We identified ten CO molecular clouds that have sizes of 5 to 29\,pc and masses of (0.8 -- 7.4) $\times 10^5\,{\rm M}_{\sun}$. They have comparable sizes and masses to the typical GMCs in our Galaxy.  
\item The HCN-to-CO flux ratio of NMA-8 is $0.024~\pm\,0.007$ and that of NMA-9 is $0.018\,\pm\,0.007$. These values are comparable to those of typical GMCs in our Galaxy and in a neighboring galaxy M31.  
\item We derived the SFE$_{\rm H\alpha,24\mu m}$ for each cloud from dust-corrected H$\alpha$ luminosity (combination of H$\alpha$ and 24~$\mu$m luminosity). We found that the SFE$_{\rm H\alpha,24\mu m}$ was in good agreement with that from 89\,GHz continuum.
\item Three CO clouds show a good spatial and velocity correlation with the bright H$\alpha$ shells. 
The CO clouds show variations in SFEs, as estimated from the 89-GHz and combination of H$\alpha$ and Spitzer 24\,$\mu$m luminosity. These variations of their properties could be explained as a snapshot of molecular clouds at different evolutionary stages.
\item A clear radial dependence of SFEs from the center of the main cavity was found. Besides, we found that the HCN and 89 \,GHz continuum emission show offset from the CO peak and are distributed to the direction of the central cluster. We interpret our results as a scenario of sequential star formation in which the massive star formation propagates through the expansion of the H$\alpha$ emission nebula excited by the central OB star cluster.
\end{enumerate}

We thank the referees and Vila Vilardo for the constructive comments that have helped to improve this manuscript. We are grateful to the staff of the Nobeyama Radio Observatory (NRO) for helping us with the data reduction. We would also like to acknowledge the excellent work by the staff working for the NMA that was shut down for common-use observations in 2007. The NRO is a branch of the National Astronomical Observatory of Japan (NAOJ), the National Institutes of Natural Sciences (NINS). This research was supported, in part, by a grant from the Hayakawa Satio Fund awarded by the Astronomical Society of Japan.





\begin{deluxetable}{lcccccccccccc}
\tabletypesize{\scriptsize}
\center
\tablecaption{$^{12}$CO($J$=1--0) Observational Properties \label{tb:fovsummary}}
\tablewidth{0pt}
\tablehead{FoV & Field Center & Array & Integration & Beam Size & P.A.& $\sigma_{\rm rms}$\tablenotemark{a}&Flux Ratio\tablenotemark{b}\\
& (RA,DEC) & Configuration & Time (hours) & (arcsec)&(degree)&(MJy sr$^{-1}$)&}
\startdata
FOV1&($01^{\rm h}34^{\rm m}31\fs7$, $+30\arcdeg47\arcmin41\farcs0$)& C,D & 8.2 & 5.4$\times$4.5&$-38.1$&75 &0.51$\pm 0.09$\\
FOV2&($01^{\rm h}34^{\rm m}32\fs7$, $+30\arcdeg47\arcmin13\farcs9$)& C,D & 7.8 & 5.4$\times$4.5&$-$38.1& 75&0.23$\pm 0.04$\\
FOV3&($01^{\rm h}34^{\rm m}33\fs3$, $+30\arcdeg46\arcmin47\farcs4$)& C,D & 6.0 & 4.1$\times$2.6&$-$46.7& 147&0.62$\pm 0.11$\\
FOV4&($01^{\rm h}34^{\rm m}30\fs9$, $+30\arcdeg46\arcmin53\farcs7$)& C,D & 7.9 & 5.4$\times$4.5&$-$38.1& 75&0.32$\pm 0.06$\\
FOV5&($01^{\rm h}34^{\rm m}34\fs5$, $+30\arcdeg46\arcmin18\farcs4$)& C,D & 6.6 & 4.1$\times$2.6&$-$46.7& 147&0.63$\pm 0.11$\\
\enddata
\tablenotetext{a}{The rms noise was measured in the emission-free region with 1-MHz resolution on the primary-beam-corrected images. The unit was converted from Jy\,beam$^{-1}$ to Jy sr$^{-1}$ using $\Omega_{\rm beam}=\frac{\pi}{4\,\rm ln\,2} {\sin} \theta_{\rm maj}{\sin} \theta_{\rm min}$, where $\Omega_{\rm beam}$ is the synthesized beam in steradian, $\theta_{\rm maj}$ and $\theta_{\rm min}$ are the synthesized beam in arcsec. The conversion factors are 6.5$\times10^{-10}$\,sr\,beam$^{-1}$ for the $5\farcs4\times4\farcs5$ resolution data, and 2.8$\times10^{-10}$\,sr\,beam$^{-1}$ for the $4\farcs1\times2\farcs6$ resolution data.}
\tablenotetext{b}{The ratio of interferometric flux to the single dish flux (R. Miura et al., 2010, in preparation), which is measured in the box region with 42\arcsec$\times$42\arcsec at the center of each FoV.}
\end{deluxetable}

\begin{deluxetable}{lccc}
\tabletypesize{\scriptsize}
\tablecaption{Observational parameters of three probes \label{tb:nmaobservationalparameter}}
\tablewidth{0pt}
\tablehead{Parameter & $^{12}$CO(1--0) & HCN(1--0) & 89\,GHz Continuum}
\startdata
Frequency (GHz) &  115.271202 & 88.6304157 & 88.9790432\,$\pm$\,0.512 (LSB)\tablenotemark{a}\\
Observing Period & 2004 Nov - 2006 Dec & 2004 Nov - 2006 Jan& 2004 Nov -
 2006 Jan \\
Observed Field &FOV1, 2, 3, 4, 5&FOV3, FOV5&FOV3, FOV5 \\
Array Configurations & C, D & C, D& C, D \\
Correlator & FX & FX &  UWBC \\
Bandwidth (MHz) & 32 & 32 & 1024 \\
FWHM of FoV (\arcsec) & 59 & 77 & 76  \\
Beam Size (\arcsec) & (see Table\,1) & 6.6$\times$5.3 & 6.9$\times$5.5 \\
Velocity Resolution (km s$ ^{-1} $) & 2.6 & 3.4 & - \\
Integration Time (hour)  & (see Table\,1) & 19 & 19\\
$\sigma_{\rm rms}$ (MJy sr$ ^{-1} $)\tablenotemark{b} & (see Table\,1) & 21 & 0.6\\
\enddata
\tablenotetext{a}{The frequency range corresponding to the line emission as HCN and HCO+(89.188526 GHz) were removed from the original visibility data.}
\tablenotetext{b}{The conversion factors are 9.3$\times10^{-10}$\,sr\,beam$^{-1}$ for the $6\farcs6\times5\farcs3$ resolution data, and 1.0$\times10^{-9}$sr\,beam$^{-1}$ for the $6\farcs9\times5\farcs5$ resolution data.}
\end{deluxetable}
\begin{deluxetable}{lcccccccccc}
\tabletypesize{\scriptsize}
\tablewidth{0pt}
\tablecaption{Cloud Properties for Clouds Found in M~33\label{tb:ngc604cloudproperties}}
\tablehead{Cloud & Alternative\tablenotemark{a}& $\alpha(J2000)$, $\delta(J2000)$\tablenotemark{b}&
 V$_{\rm LSR}$\tablenotemark{c}& $V_{\rm FWHM}$\tablenotemark{d} & $S_{\rm CO}$ \tablenotemark{e}& $\bar{D}$ \tablenotemark{f}&$M_{\rm mol}$ & $M_{\rm vir}$ \\
   No. &Name&  & (${\rm km\,s}^{-1}$) & (${\rm km\,s}^{-1}$)&(${\rm Jy\,km\,s}^{-1}$)&(pc)& (10$^5 M_{\sun}$)& (10$^5 M_{\sun}$) }
\startdata
NMA-1 	&	 	&	$01^{\rm h}34^{\rm m}31\fs59$, $+30\arcdeg46\arcmin54\farcs88$ 	&	 $-$241 	&	4.9	&	    14.4$\pm 1.5$ 	&	$<9.4$\tablenotemark{f}	&	1.6	&	$<0.3$\\
NMA-2 	&	 EPRB-101 	&	$01^{\rm h}34^{\rm m}31\fs59$, $+30\arcdeg47\arcmin42\farcs03$ 	&	 $-$244 	&	10.4	&	    22.3$\pm 1.6$ 	&	19.2	&	2.5	&	2.9\\
NMA-3 	&	 	&	$01^{\rm h}34^{\rm m}31\fs94$, $+30\arcdeg47\arcmin25\farcs14$ 	&	 $-$250 	&	7.3	&	     6.6$\pm 1.0$ 	&	$<4.9$\tablenotemark{f}	&	0.8	&	$<0.4$\\
NMA-4 	&	 	&	$01^{\rm h}34^{\rm m}32\fs29$, $+30\arcdeg46\arcmin57\farcs40$ 	&	 $-$247 	&	8.4	&	     6.8$\pm 1.0$ 	&	10.1	&	0.8	&	1.0\\
NMA-5 	&	 	&	$01^{\rm h}34^{\rm m}32\fs74$, $+30\arcdeg47\arcmin39\farcs18$ 	&	 $-$242	&	7.0	&	    10.7$\pm 1.7$ 	&	$<10.9$\tablenotemark{f}	&	1.2	&	$<0.9$\\
NMA-6 	&	 	&	$01^{\rm h}34^{\rm m}32\fs79$, $+30\arcdeg46\arcmin13\farcs55$ 	&	 $-$240 	&	7.9	&	    18.4$\pm 2.2$ 	&	21.9	&	2.1	&	1.8\\
NMA-7 	&	 WS-1	&	$01^{\rm h}34^{\rm m}33\fs25$, $+30\arcdeg46\arcmin30\farcs19$ 	&	 $-$249 	&	7.6	&	    18.6$\pm 1.7$ 	&	19.2	&	2.1	&	1.5\\
NMA-8 	&	 WS-2, EPRB-8	&	$01^{\rm h}34^{\rm m}33\fs42$, $+30\arcdeg46\arcmin47\farcs03$ 	&	 $-$243 	&	10.2	&	    65.2$\pm 2.7$ 	&	28.9	&	7.4	&	 4.1\\
NMA-9 	&	 WS-4, EPRB-9	&	$01^{\rm h}34^{\rm m}34\fs47$, $+30\arcdeg46\arcmin20\farcs00$ 	&	 $-$221 	&	9.6	&	    56.7$\pm 2.4$ 	&	27.9	&	6.4	&	   3.6\\
NMA-10 	&	 	&	$01^{\rm h}34^{\rm m}35\fs16$, $+30\arcdeg47\arcmin01\farcs50$	&	 $-$228 	&	4.8	&	    27.2$\pm 3.2$ 	&	11.0	&	3.1	&	0.4\\

\hline
NMA-11 & &$01^{\rm h}34^{\rm m}34\fs07$, $+30\arcdeg47\arcmin09\farcs47$	 & $-$239 &      $<$2.6 &    12.4$\pm 1.4$&       $<20$&    1.4 &     $<$0.7\\
NMA-12 & &$01^{\rm h}34^{\rm m}34\fs64$, $+30\arcdeg46\arcmin35\farcs10$	 & $-$228 &      $<$2.6 &    10.7$\pm 1.2$ &      $<20$&    1.2 &     $<$0.7\\
\enddata
\tablecomments{The derived properties of identified clouds were described shortly in text. Approximate errors are $\delta V_{\rm FWHM}=1.3\,{\rm km\,s}^{-1}$ (i.e., one-half the velocity resolution), $\delta M_{\rm mol}=38\,\%$, and $\delta M _{\rm vir}=40\,\%$. }
\tablenotetext{a}{WS-{\it n}: from \citet{1992ApJ...385..512W}. EPRB-{\it n}: from \citet{2003ApJS..149..343E}.}
\tablenotetext{b}{The positions of clouds were obtained from the position of the peak of emission. }
\tablenotetext{c}{The positions along the velocity axis were obtained from the position of the peak of the observed (Gaussian-fit) LSR velocities.}
\tablenotetext{d}{The full width at half-maximum velocities are deconvolved with a velocity resolution of $\delta V_{\rm FWHM}=1.3$\,km\,s$^{-1}$.}
\tablenotetext{e}{The flux densities were corrected primary beam attenuation.}
\tablenotetext{f}{$\bar{D}=D_{\alpha}+D_{\delta}$, where $D_{\alpha}$ and $D_{\delta}$ are the deconvolved full width at half-maximum diameters in right ascension and declination directions. Typical error of the cloud size differs depending on where the clouds are located because the synthesized beam size differs on each FoV. Typical errors are $\delta \bar{D}=5.1\,{\rm pc}$ for FOV3 and FOV5, $\delta \bar{D}=6.6\,{\rm pc}$ for FOV1, FOV2 and FOV4.}
\tablenotetext{g}{Since the cloud size in right ascension of NMA-1, NMA-3, and NMA-5 are less than beam size in the same direction, the cloud size $\bar{D}$ is given as the upper limit.}
\end{deluxetable}
\begin{deluxetable}{lrccccccccccccc}
\tabletypesize{\scriptsize}
\tablecaption{Cloud Properties and Star Formation Activities for Clouds\label{tb:ngc604cloudproperties2}}
\tablewidth{0pt}
\tablehead{ Cloud & $d_{\rm c}$\tablenotemark{a}& $S_{\rm HCN}$\tablenotemark{b}& $S_{\rm HCN}/S^\ast_{\rm CO}$\tablenotemark{c} & $S_{\rm 89GHz}$\tablenotemark{b} & SFE$_{\rm 89GHz}$\tablenotemark{d} & log$L_{\rm corr}({H\alpha}$)\tablenotemark{e} & SFE$_{\rm H\alpha,24{\mu}m}$\tablenotemark{h} 
\\
   No. &(pc)& (${\rm Jy\,km\,s}^{-1}$)&($10^{-2}$) &(mJy)&(\%)&(erg\,s$^{-1}$)& (\%)}
\startdata
NMA-1 & 86 &\nodata &\nodata &$<$2.6  &$<$5.0 &38.58 & 6.2$\pm$3.3\\
NMA-2 & 214&\nodata &\nodata  &\nodata &\nodata &37.53 &0.4$\pm$0.2\\
NMA-3 & 120 &\nodata &\nodata &\nodata  &\nodata &37.78 &2.5$\pm$1.3\\
NMA-4 & 35 &\nodata &\nodata & 3.4$\pm\,0.9$ &12.4$\pm\,6.6$ & 38.75 &17.2$\pm$9.3\\
NMA-5 &152 &\nodata &\nodata &\nodata  &\nodata &37.64 &1.1$\pm$0.6\\
NMA-6 &246 &$<$0.3 &$<$2.8  &$<$2.1  &$<$3.1        &37.32&0.3$\pm$0.2\\
NMA-7 &194 &$<$0.3 &$<$2.7  &$<$1.8  &$<$3.6        &37.96&$1.6\pm$0.8\\
NMA-8 & 138 & 1.1$\pm\,0.3$ &2.4$\pm\,0.7$ & 9.1$\pm\,1.5$ &5.1$\pm\,2.3$&39.08& 5.9$\pm$3.0\\
NMA-9 &320 & 0.6$\pm\,0.2$  &1.8$\pm\,0.7$ &$<$1.8 &$<$1.2&37.67 &0.3$\pm$0.1\\
NMA-10&261 &$<$0.4 &$<$2.8   &$<$2.3  &$<$2.3        &38.08 &1.1$\pm$0.6\\
\hline
NMA-1,4& 60 &$<$0.4 &$<$3.2 & \nodata  &\nodata  &\nodata &\nodata\\
\enddata
\tablenotetext{a}{The projected distance from the central cluster, $d_{\rm c}$, is given in parsec relative to the position of ($01^{\rm h}34^{\rm m}32\fs4$, $+30\arcdeg47\arcmin00\farcs3)$}
\tablenotetext{b}{For the clouds where no HCN emission was detected, the upper limits of line ratio are given by using the detection limit of $3\,\sigma_{\rm rms}$.}
\tablenotetext{c}{The CO flux densities, $S^\ast_{\rm CO}$, were measured over the extent of CO emissions ($>\,3\,\sigma_{\rm rms}$) on the convolved images with HCN synthesized beam.}
\tablenotetext{d}{The SFE$_{\rm 89\,GHz}$ was obtained using the following equation, ${\rm SFE_{\rm 89\,GHz}}=M_{\rm star}/(M_{\rm star}+M^{\ast}_{\rm mol})$, where $M^{\ast}_{\rm mol}$ represents the missing-flux-corrected molecular masses. See details in the text.}
\tablenotetext{e}{The extinction-corrected H$\alpha$ luminosity is defined as a linear combination of these two luminosities: $L_{\rm corr}({\rm H}\alpha)=L_{\rm obs}({\rm H}\alpha)+a\times L({\rm 24\,\mu m})$, where $a=(0.031\pm0.006)$ \citep{2007ApJ...666..870C}. $\delta L_{\rm corr}({\rm H}\alpha)\sim 28\,\%$.}
\tablenotetext{h}{The SFE$_{\rm H\alpha,24\mu m}$ was obtained using the same equation as SFE$_{\rm 89\,GHz}$. See details in the text.}
\end{deluxetable}
\begin{deluxetable}{lrccccccccccccc}
\tabletypesize{\scriptsize}
\tablecaption{Star Formation Efficiencies of nearby H\,{\sc ii} regions\label{tb:SFE}}
\tablewidth{0pt}
\tablehead{ Cloud & Mass (10$^5$\,M$_{\sun}$) & Q$^{\ast}$ (10$^{49}$\,s$^{-1}$)& SFE& References}
\startdata
Orion~A  &1  &  2.7\tablenotemark{a}  & 2.2    & 1\\
W49A     &7& 182   & 18.2   & 1, 2, 3\\
30~Doradus &9\tablenotemark{b} & 1122\tablenotemark{c}    & 52.8   & 2, 4 \\
\hline
NGC604 &51\tablenotemark{d} & 478\tablenotemark{e}    & 7.4   & this work\\
...NMA-4 &2\tablenotemark{d}  &28 & 12.5 & this work\\
...NMA-8 &12\tablenotemark{d}  &75 & 5.1  & this work
\enddata
\tablerefs{(1)\citet{1997ApJ...476..166W} and references therein; (2)\citet{1984ApJ...287..116K}; (3)\citet{1990MNRAS.244..458W}; (4)\citet{2008ApJS..178...56F}} 
\tablecomments{The SFE was calculated assuming the same IMF (2.35) throughout all regions. The typical error of SFE is $\sim$50\%.}
\tablenotetext{a}{The ionizing photon rates in Orion~A is determined from the stellar types in \citet{1988AJ.....95..516C}.}
\tablenotetext{b}{This value is summation of molecular cloud mass of GMCs which are annotated by 183, 186, and 189 in \citet{2008ApJS..178...56F}, which are located within the extent of H$\alpha$ emission in 30~Doradus \citep[$\sim$370\,pc;][]{1984ApJ...287..116K}. }
\tablenotetext{c}{The ionizing photon rate $Q^{\ast}$ is derived only from H$\alpha$ luminosities \citep{1984ApJ...287..116K}, and hence the SFE can give lower limit. }
\tablenotetext{d}{They represent the missing-flux-corrected molecular masses ($M^{\ast}_{\rm mol}$).}
\tablenotetext{e}{The ionizing photon rate $Q^{\ast}$ is derived from the extinction-corrected H$\alpha$ luminosities measured over a whole extent of H$\alpha$.}
\end{deluxetable}


\begin{thebibliography}{}



\bibitem[Bosch et al.(2002)]{2002MNRAS.329..481B} Bosch, G., Terlevich, E., 
\& Terlevich, R.\ 2002, \mnras, 329, 481 


\bibitem[Brinks 
\& Bajaja(1986)]{1986A&A...169...14B} Brinks, E., \& Bajaja, E.\ 1986, \aap, 169, 14 


\bibitem[Brouillet et 
al.(2005)]{2005A&A...429..153B} Brouillet, N., Muller, S., Herpin, F., Braine, J., \& Jacq, T.\ 2005, \aap, 429, 153 

\bibitem[Bruhweiler et al.(2003)]{2003AJ....125.3082B} Bruhweiler, F.~C., Miskey, C.~L., \& Smith Neubig, M.\ 2003, \aj, 125, 3082 

\bibitem[Buczilowski(1988)]{1988A&A...205...29B} Buczilowski, U.~R.\ 1988, \aap, 205, 29 

\bibitem[Calzetti et al.(2007)]{2007ApJ...666..870C} Calzetti, D., et al.\ 
2007, \apj, 666, 870
\bibitem[Cardelli 
\& Clayton(1988)]{1988AJ.....95..516C} Cardelli, J.~A., \& Clayton, G.~C.\ 1988, \aj, 95, 516 






\bibitem[Churchwell 
\& Goss(1999)]{1999ApJ...514..188C} Churchwell, E., \& Goss, W.~M.\ 1999, \apj, 514, 188 

\bibitem[Conti 
\& Blum(2002)]{2002ApJ...564..827C} Conti, P.~S., \& Blum, R.~D.\ 2002, \apj, 564, 827 


\bibitem[Corbelli 
\& Salucci(2000)]{2000MNRAS.311..441C} Corbelli, E., \& Salucci, P.\ 2000, \mnras, 311, 441 


\bibitem[Deharveng et 
al.(2005)]{2005A&A...433..565D} Deharveng, L., Zavagno, A., \& Caplan, J.\ 2005, \aap, 433, 565 
\bibitem[Deharveng et 
al.(2008)]{2008A&A...482..585D} Deharveng, L., Lefloch, B., Kurtz, S., Nadeau, D., Pomar{\`e}s, M., Caplan, J., \& Zavagno, A.\ 2008, \aap, 482, 585 

\bibitem[Deul 
\& den Hartog(1990)]{1990A&A...229..362D} Deul, E.~R., \& den Hartog, R.~H.\ 1990, \aap, 229, 362 
\bibitem[Diaz et al.(1996)]{1996ASPC...98..399D} Diaz, A.~I., Terlevich, 
E., Terlevich, R., Gonzalez-Delgado, R.~M., Perez, E., 
\& Garcia-Vargas, M.~L.\ 1996, From Stars to Galaxies: the Impact of Stellar Physics on Galaxy Evolution, 98, 399 




\bibitem[Drissen et al.(1993)]{1993AJ....105.1400D} Drissen, L., Moffat, 
A.~F.~J., \& Shara, M.~M.\ 1993, \aj, 105, 1400 


\bibitem[Elmegreen 
\& Lada(1977)]{1977ApJ...214..725E} Elmegreen, B.~G., \& Lada, C.~J.\ 1977, \apj, 214, 725 

\bibitem[Elmegreen(1998)]{1998ASPC..148..150E} Elmegreen, B.~G.\ 1998, 
Origins, 148, 150 



\bibitem[Engargiola et al.(2003)]{2003ApJS..149..343E} Engargiola, G., 
Plambeck, R.~L., Rosolowsky, E., \& Blitz, L.\ 2003, \apjs, 149, 343 

\bibitem[Fari{\~n}a et al.(2010)]{2010IAUS..266..391F} Fari{\~n}a, C., 
Bosch, G.~L., \& Barb{\'a}, R.~R.\ 2010, IAU Symposium, 266, 391 


\bibitem[Freedman et al.(2001)]{2001ApJ...553...47F} Freedman, W.~L., et 
al.\ 2001, \apj, 553, 47 

\bibitem[Fukui et al.(2008)]{2008ApJS..178...56F} Fukui, Y., et al.\ 2008, 
\apjs, 178, 56 



\bibitem[Gonz{\'a}lez Delgado 
\& P{\'e}rez(2000)]{2000MNRAS.317...64G} Gonz{\'a}lez Delgado, R.~M., \& P{\'e}rez, E.\ 2000, \mnras, 317, 64 


\bibitem[Heiner et al.(2009)]{2009ApJ...700..545H} Heiner, J.~S., Allen, 
R.~J., \& van der Kruit, P.~C.\ 2009, \apj, 700, 545 

\bibitem[Helfer 
\& Blitz(1997a)]{1997ApJ...478..162H} Helfer, T.~T., \& Blitz, L.\ 1997a, \apj, 478, 162 

\bibitem[Helfer 
\& Blitz(1997b)]{1997ApJ...478..233H} Helfer, T.~T., \& Blitz, L.\ 1997b, \apj, 478, 233 

\bibitem[Hoopes \& Walterbos(2000)]{2000ApJ...541..597H} Hoopes, C.~G., \& Walterbos, R.~A.~M.\ 2000, \apj, 541, 597 


\bibitem[Hoopes et al.(2001)]{2001ApJ...559..878H} Hoopes, C.~G., 
Walterbos, R.~A.~M., \& Bothun, G.~D.\ 2001, \apj, 559, 878 

\bibitem[Hunter et al.(1996)]{1996ApJ...456..174H} Hunter, D.~A., Baum, 
W.~A., O'Neil, E.~J., Jr., \& Lynds, R.\ 1996, \apj, 456, 174 

\bibitem[Kennicutt(1984)]{1984ApJ...287..116K} Kennicutt, R.~C., Jr.\ 1984, 
\apj, 287, 116 

\bibitem[Kennicutt et al.(2007)]{2007ApJ...671..333K} Kennicutt, R.~C., 
Jr., et al.\ 2007, \apj, 671, 333 






\bibitem[Ma{\'{\i}}z-Apell{\'a}niz et al.(2004)]{2004AJ....128.1196M} 
Ma{\'{\i}}z-Apell{\'a}niz, J., P{\'e}rez, E., 
\& Mas-Hesse, J.~M.\ 2004, \aj, 128, 1196 


\bibitem[Martins et 
al.(2005)]{2005A&A...436.1049M} Martins, F., Schaerer, D., \& Hillier, D.~J.\ 2005, \aap, 436, 1049 
\bibitem[Mayya(1994)]{1994AJ....108.1276M} Mayya, Y.~D.\ 1994, \aj, 108, 
1276 
\bibitem[Mayya(1995)]{1995AJ....109.2503M} Mayya, Y.~D.\ 1995, \aj, 109, 
2503 
\bibitem[Mayya 
\& Prabhu(1996)]{1996AJ....111.1252M} Mayya, Y.~D., \& Prabhu, T.~P.\ 1996, \aj, 111, 1252 

\bibitem[Melnick(1980)]{1980A&A....86..304M} Melnick, J.\ 1980, \aap, 86, 304 

\bibitem[Muraoka et al.(2007)]{2007PASJ...59...43M} Muraoka, K., et al.\ 
2007, \pasj, 59, 43 




\bibitem[Okumura et al.(2000)]{2000PASJ...52..393O} Okumura, S.~K., et al.\ 
2000, \pasj, 52, 393 


\bibitem[Rand 
\& Kulkarni(1990)]{1990ApJ...349L..43R} Rand, R.~J., \& Kulkarni, S.~R.\ 1990, \apjl, 349, L43 

\bibitem[Rela{\~n}o 
\& Kennicutt(2009)]{2009ApJ...699.1125R} Rela{\~n}o, M., \& Kennicutt, R.~C.\ 2009, \apj, 699, 1125 

\bibitem[Rieke et al.(2004)]{2004ApJS..154...25R} Rieke, G.~H., et al.\ 
2004, \apjs, 154, 25 

\bibitem[Rosolowsky et al.(2003)]{2003ApJ...599..258R} Rosolowsky, E., 
Engargiola, G., Plambeck, R., \& Blitz, L.\ 2003, \apj, 599, 258 



\bibitem[Scoville 
\& Sanders(1987)]{1987ASSL..134...21S} Scoville, N.~Z., \& Sanders, D.~B.\ 1987, Interstellar Processes, 134, 21 

\bibitem[Scoville et al.(1991)]{1991ApJ...366L...5S} Scoville, N.~Z., 
Sargent, A.~I., Sanders, D.~B., \& Soifer, B.~T.\ 1991, \apjl, 366, L5 


\bibitem[Solomon et al.(1992)]{1992ApJ...387L..55S} Solomon, P.~M., Downes, 
D., \& Radford, S.~J.~E.\ 1992, \apjl, 387, L55 


\bibitem[Strong et 
al.(1988)]{1988A&A...207....1S} Strong, A.~W., et al.\ 1988, \aap, 207, 1 









\bibitem[Tenorio-Tagle et al.(2000)]{2000ApJ...541..720T} Tenorio-Tagle, 
G., Mu{\~n}oz-Tu{\~n}{\'o}n, C., P{\'e}rez, E., Ma{\'{\i}}z-Apell{\'a}niz, 
J., \& Medina-Tanco, G.\ 2000, \apj, 541, 720 



\bibitem[Tosaki et al.(2007)]{2007ApJ...664L..27T} Tosaki, T., Miura, R., 
Sawada, T., Kuno, N., Nakanishi, K., Kohno, K., Okumura, S.~K., 
\& Kawabe, R.\ 2007, \apjl, 664, L27 


\bibitem[Tsutsumi et al.(1997)]{1997ASPC..125...50T} Tsutsumi, T., Morita, 
K.-I., 
\& Umeyama, S.\ 1997, Astronomical Data Analysis Software and Systems VI, 125, 50 

\bibitem[T{\"u}llmann et al.(2008)]{2008ApJ...685..919T} T{\"u}llmann, R., 
et al.\ 2008, \apj, 685, 919 



\bibitem[Viallefond et 
al.(1992)]{1992A&A...265..437V} Viallefond, F., Boulanger, F., Cox, P., Lequeux, J., Perault, M., \& Vogel, S.~N.\ 1992, \aap, 265, 437 





\bibitem[Ward-Thompson 
\& Robson(1990)]{1990MNRAS.244..458W} Ward-Thompson, D., \& Robson, E.~I.\ 1990, \mnras, 244, 458 

\bibitem[Werner et al.(2004)]{2004ApJS..154....1W} Werner, M.~W., et al.\ 
2004, \apjs, 154, 1 
\bibitem[Whitworth et al.(1994)]{1994MNRAS.268..291W} Whitworth, A.~P., 
Bhattal, A.~S., Chapman, S.~J., Disney, M.~J., 
\& Turner, J.~A.\ 1994, \mnras, 268, 291 



\bibitem[Williams 
\& McKee(1997)]{1997ApJ...476..166W} Williams, J.~P., \& McKee, C.~F.\ 1997, \apj, 476, 166 

\bibitem[Wilson 
\& Scoville(1990)]{1990ApJ...363..435W} Wilson, C.~D., \& Scoville, N.\ 1990, \apj, 363, 435 

\bibitem[Wilson 
\& Scoville(1992)]{1992ApJ...385..512W} Wilson, C.~D., \& Scoville, N.\ 1992, \apj, 385, 512 



\bibitem[Yang et al.(1996)]{1996AJ....112..146Y} Yang, H., Chu, Y.-H., 
Skillman, E.~D., \& Terlevich, R.\ 1996, \aj, 112, 146 
\bibitem[Zavagno et 
al.(2006)]{2006A&A...446..171Z} Zavagno, A., Deharveng, L., Comer{\'o}n, F., Brand, J., Massi, F., Caplan, J., \& Russeil, D.\ 2006, \aap, 446, 171 


\end{thebibliography}
\end{document}